%% file: belle-ds2lnu-ds2hadronic-brs.tex
\documentclass[12pt]{article}
\usepackage{graphicx}

\newcommand\pubnumber{}
\newcommand\pubdate{\today}

\newcommand{\dtagkx}{D_{\rm tag}KX_{\rm frag}}
\newcommand{\ds}{D_{s}}

\newcommand{\dsst}{D^{\ast}_{s}}
\newcommand{\xfrag}{X_{\rm frag}}
\newcommand{\dssttodsg}{D^{\ast}_{s}\to\ds\gamma}

\newcommand{\dtag}{D_{\rm tag}}

\newcommand{\dz}{D^0}
\newcommand{\dc}{D^+}
\newcommand{\lc}{\Lambda_c^+}
\newcommand{\dsmunu}{\ds^+\to\mu^+\nu_{\mu}}

\newcommand{\br}{{\cal B}}
\newcommand{\fds}{f_{\ds}}
\newcommand{\fb}{fb$^{-1}$}

\newcommand{\munu}{\mu\nu_{\mu}}
\newcommand{\taunu}{\tau\nu_{\tau}}
\newcommand{\taumunu}{\tau(\mu)\nu_{\tau}}
\newcommand{\tauenu}{\tau(e)\nu_{\tau}}
\newcommand{\taupinu}{\tau(\pi)\nu_{\tau}}

\newcommand{\mmiss}{M_{\rm miss}}
\newcommand{\fbias}{f_{\rm bias}}
\newcommand{\eecl}{E_{\rm ECL}}
\newcommand{\etal}{{\it et al.}}

\def\kit{Karlsruher Institut f\"ur Technologie\\
Institut f\"ur Experimentelle Kernphysik, 76131 Karlsruhe, Germany}
\def\support{For the Belle Collaboration}

\def\Title#1{\begin{center} {\Large #1 } \end{center}}
\def\Author#1{\begin{center}{ \sc #1} \end{center}}
\def\Address#1{\begin{center}{ \it #1} \end{center}}

\newcommand\pubblock{\rightline{\begin{tabular}{l} \pubnumber\\
         \pubdate  \end{tabular}}}
\newenvironment{Abstract}{\begin{quotation}  }{\end{quotation}}
\newenvironment{Presented}{\begin{quotation} \begin{center} 
             PRESENTED AT\end{center}\bigskip 
      \begin{center}\begin{large}}{\end{large}\end{center} \end{quotation}}
\def\Acknowledgements{\bigskip  \bigskip \begin{center} \begin{large}
             \bf ACKNOWLEDGEMENTS \end{large}\end{center}}
\input econfmacros.tex
\begin{document}
\begin{titlepage}
\pubblock

\vfill
\Title{Improved measurements of $D_s$ meson decay constant and branching fractions of $\ds^+\to K^-K^+\pi^+$, $\overline{K}{}^0K^+$ and $\eta\pi^+$ decays from Belle}
\vfill
\Author{ An\v ze Zupanc\\\support}
\Address{\kit}
\vfill
\begin{Abstract}
We present preliminary absolute branching fraction measurements of leptonic
$\ds^+$ decays to $\mu^+\nu_{\mu}$ and $\tau^+\nu_{\tau}$, and of hadronic $\ds^+$ decays to $K^-K^+\pi^+$, $\overline{K}{}^0K^+$ and $\eta\pi^+$. 
The results are obtained from a large data sample 
collected near the $\Upsilon(4S)$ and $\Upsilon(5S)$ resonances with the Belle detector 
at the KEKB asymmetric-energy $e^+e^-$ collider. We obtain the the branching fractions
$\br(\ds^+\to\mu^+\nu_{\mu})=(0.528\pm0.028\pm0.019)\%$ and $\br(\ds^+\to\tau^+\nu_{\tau})=(5.70\pm0.21{}^{+0.31}_{-0.30})\%$ which 
are combined to determine the $\ds$ decay constant $\fds=(255.0\pm4.2\pm5.0)$~MeV, where the first and second uncertainties
are statistical and systematic, respectively. Branching fractions of hadronic decays are measured to be
$\br(\ds^+\to K^-K^+\pi^+)=(5.06 \pm 0.15\pm 0.19)\%$, $\br(\ds^+\to\overline{K}{}^0K^+)=(2.84 \pm 0.12\pm 0.08)\%$ and $\br(\ds^+\to\eta\pi^+)=(1.79\pm0.14\pm 0.05)\%$.
\end{Abstract}
\vfill
\begin{Presented}
The 5th International Workshop on Charm Physics\\
Honolulu, Hawai'i, USA, May 14--17, 2012
\end{Presented}
\vfill
\end{titlepage}
\def\thefootnote{\fnsymbol{footnote}}
\setcounter{footnote}{0}

\section{Introduction}

The leptonic decays of mesons provide access to experimentally clean measurements of
the meson decay constants or the relevant Cabibbo-Kobayashi-Maskawa
matrix elements. In the Standard Model (SM) the branching fraction for a leptonic decay of 
a charged pseudoscalar meson, such as $D^+_s$, is given by \cite{PDG,Rosner:2012bb}:
\begin{equation}
 \br(D_{s}^+\to \ell^+\nu_{\ell})=\frac{\tau_{\ds}M_{D_{s}}}{8\pi}f_{D_{s}}^2G_F^2|V_{cs}|^2m_{\ell}^2\left(1-\frac{m_{\ell}^2}{M_{D_{s}}^2} \right)^2,
 \label{eq:brleptonic_sm}
\end{equation}
where $M_{D_{s}}$ is the $D_{s}$ mass, $\tau_{\ds}$ is its lifetime, $m_{\ell}$ is the lepton mass, $V_{cs}$ is the 
Cabibbo-Kobayashi-Maskawa (CKM) matrix element between the $\ds$ constituent quarks $c$ and $s$, and $G_F$ is the Fermi coupling constant. 
The parameter $f_{D_{s}}$ is the decay constant, and is related to the wave-function overlap of the quark and anti-quark. 
The leptonic decays of pseudoscalar mesons are suppressed by helicity conservation and their decay rates are thus proportional
to the square of the lepton mass. Leptonic $\ds$ decays into electrons are not observable whereas decays to taus are favored over decays
to muons in spite of the reduced phase-space in the former case.

If the magnitude of the relevant CKM matrix element is well
known from other measurements then by measuring the leptonic branching fraction of a pseudoscalar meson one
can determine the decay constant with high precision. Conversely, if one can precisely
estimate the decay constant of a pseudoscalar meson, it is possible to determine the
magnitude of the relevant CKM element.

Measurements of $\fds$ have been made by several groups: CLEO-c \cite{CLEO-c,CLEO-rho,CLEO-CSP}, 
Belle \cite{Belle-munu} and BaBar \cite{Sanchez}. Rosner and Stone combined the above measurements and report the 
experimental world average to be $\fds^{\rm exp}=(260\pm5.4)$~MeV \cite{Rosner:2012bb}. Within the SM, $\fds$ has been predicted 
using several methods \cite{Lat:Dav,Lat:Milc,Lat:Nf2,Bordes,Lucha,Field,LF}. While most calculations give values lower 
than the $\fds$ measurement, the errors on predicted values are too large in most cases to claim any disagreement with experiment. The largest
discrepancy (2.0 standard deviations) is with an unquenched lattice QCD (LQCD) calculation: $\fds^{\rm LQCD}=(248\pm2.5)$~MeV \cite{Lat:Dav}.  
There are several theoretical scenarios in which non SM particles may modify the leptonic decay rates of the $\ds$ meson. 
Akeroyd and Chen pointed out that leptonic decay widths are modified in two-Higgs-doublet models (2HDM) \cite{AkeroydC}. 
Measurements of $\fds$ with an accuracy that matches the precision of theoretical calculations are 
thus necessary in order to discover or constrain effects of NP.

In these proceedings we present preliminary results of absolute branching fraction measurements of $\ds^+\to\mu^+\nu_{\mu}$ and 
$\ds^+\to\tau^+\nu_{\tau}$ decays\footnote{Charge conjugation is assumed throughout this note unless stated otherwise.} which supersede the previous measurement of $\br(\ds^+\to\mu^+\nu_{\mu})$ by Belle \cite{Belle-munu}
performed on a smaller data sample. This analysis is based on a data sample of 
$913$~fb$^{-1}$ recorded near $\sqrt{s}=10.68$ GeV by the Belle detector at the KEKB asymmetric-energy collider \cite{KEKB}. 

\section{Belle detector}
The Belle detector is a large-solid-angle magnetic spectrometer that consists of a silicon vertex detector (SVD), a 50-layer central drift chamber (CDC), an array of
aerogel threshold Cherenkov counters (ACC), a barrel-like arrangement of time-of-flight scintillation counters (TOF), and an electromagnetic calorimeter
(ECL) comprised of CsI(Tl) crystals located inside a superconducting solenoid coil that provides a 1.5~T magnetic field.  An iron flux-return located outside of
the coil is instrumented to detect $K_L^0$ mesons and to identify muons (KLM). The detector is described in detail elsewhere~\cite{Belle}.
Two inner detector configurations were used. A 2.0 cm beampipe and a 3-layer silicon vertex detector was used for the first sample
of 156 fb$^{-1}$, while a 1.5 cm beampipe, a 4-layer silicon detector and a small-cell inner drift chamber were used to record  
the remaining 757 fb$^{-1}$ of data.

Tracks are detected with the CDC and the SVD. They are required to have an impact parameter with respect to the 
interaction point of less than 0.5~cm in the radial direction and less than 1.5~cm in the beam direction. 
A likelihood ratio for a given track to be a kaon or pion, ${\cal L}_{(K, \pi)}$, is obtained by utilizing 
specific ionization energy loss measurements in the CDC, light yield measurements from the ACC, and time-of-flight 
information from the TOF. For electron identification we use position, cluster energy, shower
shape in the ECL, combined with track momentum and $dE/dx$ measurements in the CDC and hits in the ACC.
For muon identification, we extrapolate the CDC track to the KLM and compare the measured range and transverse 
deviation in the KLM with the expected values. Photons are required to have energies in the laboratory
frame of at least 50 - 100 MeV, depending on the detecting part of the ECL. Neutral pion candidates are 
reconstructed using photon pairs with invariant mass between 120 and 150~MeV\footnote{We use natural units throughout this note.}. Neutral kaon candidates are 
reconstructed using charged pion pairs with invariant mass within $\pm20$~MeV of the nominal $K^0$ mass.

\section{Overview of the method}
The method of absolute branching fraction measurement of $\ds^-\to\ell^-\overline{\nu}{}_{\ell}$ decays is similar to the one previously used
by the Belle collaboration \cite{Belle-munu} and more recently by the Babar collaboration \cite{Sanchez}.
In this method the $e^+e^- \to c\bar{c}$ events which contain $\ds^-$ mesons produced through the following reactions:
\begin{equation}
 e^+e^-\to c\bar{c}\to \dtagkx\ds^{\ast -},~\ds^{\ast -}\to\ds^-\gamma, 
 \label{eq:signal_events_type}
\end{equation}
are fully reconstructed in two steps. In these events one of the two charm quarks hadronizes into a $\ds^{\ast -}$ meson while the other quark
hadronizes into a charm hadron denoted as $\dtag$ (tagging charm hadron). We reconstruct the tagging charm
hadron as $\dz$, $\dc$, $\lc$\footnote{In events where $\lc$ is reconstructed as tagging charm hadron additional $\overline{p}$ is reconstructed
in order to conserve the total baryon number.}, $D^{\ast +}$ or $D^{\ast 0}$. %or $\Lambda_c(2595,2625)^+$. 
The strangeness of the event is conserved
by requiring additional kaon, denoted as $K$, which can be either $K^+$ or $K^0_S$. Since Belle collected data at energies well above 
$D{}^{(\ast)}_{\rm tag} K D_s^{\ast}$ threshold additional particles can be produced in the process of hadronization. These particles are 
denoted as $\xfrag$ and can be: even number of kaons and or any number of pions or photons. In this measurement only pions are considered 
when reconstructing the fragmentation system $\xfrag$. We require $\ds^-$ mesons to be produced in a
$\ds^{\ast -}\to\ds^-\gamma$ decay which provides powerful kinematic constraint 
($\dsst$ mass, or mass difference between $\dsst$ and $\ds$) that improves the resolution of the missing mass (defined below)
and suppresses the combinatorial background.

In the first step of the measurement no requirements are placed on the daughters of the signal $\ds^-$ meson
in order to obtain a fully inclusive sample of $\ds^-$ events which is used for normalization
in the calculation of the branching fractions. The number of inclusively reconstructed $\ds$ mesons is
extracted from the distribution of events in the missing mass, $\mmiss(\dtagkx\gamma)$, recoiling against the $\dtagkx\gamma$ system:
\begin{equation}
\mmiss(\dtagkx\gamma)  =  \sqrt{p_{\rm miss}(\dtagkx\gamma)^2},
\label{eq:massds} 
\end{equation}
where $p_{\rm miss}$ is the missing momentum in the event:
\begin{equation}
p_{\rm miss}(\dtagkx\gamma)  =  p_{e^+} + p_{e^-} - p_{\dtag} - p_{K} - p_{\xfrag} - p_{\gamma}.\\
\label{eq:pmiss}
\end{equation}
Here, $p_{e^+}$ and $p_{e^-}$ are the momenta of the colliding positron and electron beams, respectively, and the $p_{\dtag}$, $p_{K}$, 
$p_{\xfrag}$, and $p_{\gamma}$ are the measured momenta of the reconstructed $\dtag$, kaon, fragmentation system and the photon from
$\dssttodsg$ decay, respectively. Correctly reconstructed events given in the Eq. \ref{eq:signal_events_type} produce a peak in the 
$\mmiss(\dtagkx\gamma)$ at nominal $\ds$ meson mass.

In the second step we search for the purely leptonic $\ds^+\to\mu^+\nu_{\mu}$ and $\ds^+\to\tau^+\nu_{\tau}$ decays within the inclusive $\ds^+$ 
sample by requiring an additional charged track identified as an electron, muon or charged pion to be present in the rest of the event. 
In case of $\ds^+\to\tau^+\nu_{\tau}$ decays the electron, muon or charged pion track identifies the subsequent $\tau^+$ decay to 
$e^+\nu_e\overline{\nu}{}_{\tau}$, $\mu^+\nu_{\mu}\overline{\nu}{}{\tau}$ or $\pi^+\overline{\nu}{}_{\tau}$. Hadronic decays, $\ds^+\to\overline{K}{}^0K^+$ and $\eta\pi^+$, are reconstructed
partially by only requiring that an additional charged kaon or pion is present in the rest of the event while no requirements are set upon the 
neutral hadrons ($\overline{K}{}^0$ or $\eta$) in order to increase the reconstruction efficiency. 
In case of $\ds^+\to K^-K^+\pi^+$ all three charged tracks are required to be present in the rest of the event. 

\section{\boldmath Inclusive $\ds$ reconstruction}

The reconstruction of the inclusive $\ds$ sample starts with the reconstruction of the tagging charmed hadrons, $\dtag$. In order to 
increase the reconstruction efficiency of studied events, the $\dtag$ is reconstructed in as many decay modes as possible, while 
keeping the purity of the sample at reasonable level. The ground state $\dtag$ ($\dz$, $\dc$, $\lc$) hadrons are reconstructed in 
total in 18 hadronic decay modes (see Table \ref{tab:dtag_modes}).
Only modes with up to one $\pi^0$ in the final state are used in order to avoid large backgrounds.
\begin{table}[t]\scriptsize
\centering
  \begin{tabular}{l|c}
  $\dz\to$ 			& ${\cal B}~[\%]$\\\hline
  $K^-\pi^+$			& 3.9 \\
  $K^-\pi^+\pi^0$		& 13.9 \\
  $K^-\pi^+\pi^+\pi^-$		& 8.1 \\
  $K^-\pi^+\pi^+\pi^-\pi^0$	& 4.2 \\
  $K^0_S\pi^+\pi^-$		& 2.9 \\
  $K^0_S\pi^+\pi^-\pi^0$	& 5.4 \\ \hline
  Sum                           & 38.4
  \end{tabular}  
\hspace{0.05\textwidth}
    \begin{tabular}{l|c}
   $\dc\to$ 			& ${\cal B}~[\%]$\\\hline
   $K^-\pi^+\pi^+$		& 9.4 \\
   $K^-\pi^+\pi^+\pi^0$		& 6.1 \\
   $K^0_S\pi^+$			& 1.5 \\
   $K^0_S\pi^+\pi^0$		& 6.9 \\
   $K^0_S\pi^+\pi^+\pi^-$	& 3.1 \\
   $K^+K^-\pi^+$		& 1.0 \\ \hline
   Sum                          & 28.0
  \end{tabular}
\hspace{0.05\textwidth}
    \begin{tabular}{l|c}
   $\lc\to$ 			& ${\cal B}~[\%]$\\\hline
   $pK^-\pi^+$			& 5.0 \\
   $pK^-\pi^+\pi^0$		& 3.4 \\
   $pK^0_S$			& 1.1 \\
   $\Lambda\pi^+$			& 1.1 \\
   $\Lambda\pi^+\pi^0$		& 3.6 \\
   $\Lambda\pi^+\pi^+\pi^-$	& 2.6 \\ \hline
   Sum                          & 16.8
  \end{tabular}
\caption{Summary of $\dtag=\dz$, $\dc$ and $\lc$ decay modes used in this measurement. The branching fractions are taken
from \cite{PDG}.}
\label{tab:dtag_modes}
 \end{table}

The center of mass momentum of the $\dtag$ candidates ($p^{\ast}$) is required to be greater than 2.3 (or 2.5 for less clean $\dtag$ modes) GeV/$c$ in order to 
reduce the background and to remove charmed hadrons originating from $B$ decays. The decay products of the $\dtag$ candidates are fitted 
to a common vertex and candidates with poor fit quality are discarded ($\chi^2/n.d.f<20$). The purity of $\dtag$ sample at this stage is rather
low - around 17\% in the signal region defined as $\pm3\sigma$ interval around the nominal $\dtag$ mass, where $\sigma$ 
is the $\dtag$ decay mode dependent invariant mass resolution. In order to further clean up the $\dtag$ sample we train an 
NeuroBayes~\cite{NB} neural network using small sample of real data 
(around 1\% of the total sample). Network combines information from the following input variables into one single scalar output variable:
the distance between the decay and the production vertices of $\dtag$ candidates in $r-\phi$ plane,
the $\chi^2/n.d.f$ of the vertex fit of $\dtag$ candidates, the cosine of the angle between the $\dtag$ momentum vector and the vector joining
its decay and production vertices in $r-\phi$ plane, the cosine of the angle between the momentum of one of the $\dtag$ daughters momentum
vector in the $\dtag$ rest frame and the $\dtag$ momentum in the laboratory frame (only for two-body $\dtag$ decays), particle identification
likelihood ratios and for the $\dtag$ decay modes including $\pi^0$ the minimal energy of the two photons. To obtain the signal and background
distributions of variables entering the NeuroBayes network a statistical tool to unfold the data distributions (sPlot) is 
applied \cite{Pivk:2004ty}. The cut on the network output variable is optimized for each $\dtag$ mode individually by maximizing $S/\sqrt{S+B}$,
where $S$ ($B$) refers to the signal (background) yield in the signal window of $\dtag$ invariant mass determined by performing a fit to the
$\dtag$ invariant mass distribution. After the optimization the purity of the correctly reconstructed $\dtag$ candidates 
increases from 17\% to 42\% while loosing only 
around 16\% of signal $\dtag$ candidates. We keep only $\dtag$ candidates from signal region in $\dtag$ invariant mass in rest of the analysis.

Once the ground state $\dtag$ hadrons have been reconstructed, $\dz$ and $\dc$ mesons originating from
$D^{\ast}$ decays are identified by reconstructing the decays $D^{\ast +}\to \dz\pi^+$, $\dc\pi^0$ and $D^{\ast 0}\to \dz\pi^0$, $\dz\gamma$. 
The motivation for this reconstruction is to clean up the subsequent $K\xfrag\gamma$ reconstruction; by absorbing one more particle to the 
tagging charm hadron the subsequent combinatoric background can be reduced. In addition, by reconstructing 
$D^{\ast +}\to \dz\pi^+$ decays we can determine of the charm quantum number of $\dz$ decays including neutral kaons.
The slow pions from $D^{\ast}$ decay are refitted to a $D$ production vertex in order to improve the resolution of the mass difference, 
$\Delta M = M(D\pi)-M(D)$. The photon energy in the laboratory frame is required to be larger than 50 MeV for $\pi^0\to\gamma\gamma$ and
175 MeV for $D^{\ast 0}\to \dz\gamma$ decays. In the latter decays the $\gamma$ candidate is combined with any other photon in a event, 
and if there exist a combination of two photons with invariant mass within 10 MeV/$c^2$ around the nominal $\pi^0$ mass and their energy 
asymmetry ($(E_{\gamma_1}-E_{\gamma_2})/(E_{\gamma_1}+E_{\gamma_2})$) is smaller than 0.5 the $D^{\ast 0}$ candidate is rejected. For all
$D^{\ast}$ decays, the mass difference $m(D^{\ast})-m(D)$ is required to be within 3$\sigma$ of the corresponding nominal mass difference.

A $K$ candidate is selected to be either $K^{\pm}$ or $K^0_S$ candidates and does not 
overlap with the $\dtag$ candidate. 

From the remaining tracks and $\pi^0$ candidates in the event that do not overlap with the $\dtag K$ candidate we form $\xfrag$ candidates. 
Only modes with up to three pions and up to one $\pi^0$ are used in order to avoid large combinatoric background. In addition pions must have 
momentum larger than 100 MeV/$c$ in the laboratory frame. At this stage no requirement is applied to the total charge of the $\xfrag$ system.

The $\dtag$, $\xfrag$ and $K$ candidates are combined to form a $\dtagkx$ combinations and we keep only those with total charge $\pm 1$. The charm
and strange quark content of the $\dtagkx$ system is required to be consistent with that recoiling from a $\dsst$: if $\dtag$ is 
reconstructed in flavor specific decay mode and the primary kaon candidate is charged it is required that the kaon
charge and the charm quantum number of $\dtag$ are opposite to the $\dsst$ charge; if primary kaon candidate is a 
$K^0_S$ the charm quantum number of $\dtag$ is required to be opposite to the $\dsst$ charge; and if $\dtag$ is reconstructed in a 
self-conjugated decay mode the charge of the primary kaon is required to be opposite to the $\dsst$ charge. 
A kinematic fit to $\dtagkx$ candidate is performed in which the particles are required to originate from a common
point inside the IP region, and the $\dtag$ mass is constrained to the nominal value. We select only one $\dtagkx$ candidate in an event which has missing mass, 
$\mmiss(\dtagkx)=\sqrt{|p_{e^+} + p_{e^-} - p_{\dtag} - p_{K} - p_{\xfrag}|^2}$, is closest to the nominal $\dsst$ mass and within 
$2.00~\mbox{GeV/$c^2$}<\mmiss(\dtagkx)<2.25~\mbox{GeV/$c^2$}$ interval (corresponding to around $\pm3\sigma$ interval). 

Finally, a photon candidate is identified which is consistent with the decay $\dssttodsg$ and
does not overlap with the $\dtagkx$ system. We require that the energy of the photon candidate is larger then 120 MeV in the laboratory frame 
and that the cosine of the angle between the direction of $\dtag$ hadron and the direction of the photon candidate is negative, 
since the signal photon should be in opposite hemisphere of the event with respect to the $\dtag$. We perform similar kinematic fit
with the signal photon included and with the missing mass recoiling against the $\dtagkx$ constrained to the nominal $\dsst$ mass. 
All $\dtagkx\gamma$ candidates are required  to have $p_{\rm miss}^{\ast}(\dtagkx\gamma)>2.8$ GeV/$c$ and $\mmiss(\dtagkx\gamma)>1.83$ GeV/$c^2$ 
(see Eqs. \ref{eq:pmiss} and \ref{eq:massds}). After the final selections, there are in average 2.1 $\dtagkx\gamma$ candidates per event which
are solely due to multiple $\gamma$ candidates. We select a best $\dtagkx\gamma$ candidate to be the one with the highest NeuroBayes network output 
which is trained to separate signal photons from photons produced in other decays. A relative gain of
around 23\% in absolute reconstruction efficiency is obtained by applying the best $\dtagkx\gamma$ candidate selection instead of 
completely random selection. Figure \ref{fig:dsincl:datafitres} shows the distribution of $\mmiss(\dtagkx\gamma)$ for each $\xfrag$
mode separately. 

\subsection{\boldmath Inclusive $\ds$ yield extraction}
\label{sec:inclds:yield:extraction}
The yield of inclusively reconstructed $\ds$ mesons is determined by performing a fit to the missing mass $\mmiss(\dtagkx\gamma)$ distribution
for each $\xfrag$ mode individually. The components of the fit are divided into six categories: signal, 
mis-reconstructed signal event ($K$ candidate or one of the pions forming $\xfrag$ system candidate originate from $\ds$ decay), 
$D^{\ast0}\to D^0\gamma$ (signal $\gamma$ candidate originates from $D^{\ast0}$ decay), $D_{(s)}^{\ast}\to D_{(s)}\pi^0$ (one of the photons
from $\pi^0$ decay from $D_{(s)}^{\ast}\to D_{(s)}\pi^0$ decay is taken to be signal $\gamma$ candidate), wrong $\gamma$ (the energy deposited
in the ECL was produced by unmatched charged track or by beam induced interactions), and $\gamma$ from $\pi^0$ (signal photon originates from a $\pi^0$
decay which does not originate from $D^{\ast}_{(s)}$ decay). Each of the above six components is parameterized with a non-parametric 
histogram probability density function (PDF), ${\cal H}(\mmiss(\dtagkx\gamma))$, taken from a large sample of Monte Carlo (MC) simulated events. The total PDF
for given $\xfrag$ mode is thus given by:
\begin{eqnarray}
 {\cal F}^{\xfrag}(\mmiss(\dtagkx\gamma)) & = & N_{\rm sig}{\cal H}_{\rm sig}^{\xfrag}(\mmiss(\dtagkx\gamma))\otimes {\cal G}(\sigma_{\rm cal}) \nonumber\\
 & & + \sum_{i=1}^5 N_{i-{\rm bkg}}^{\xfrag}{\cal H}_{i-{\rm bkg}}^{\xfrag}(\mmiss(\dtagkx\gamma)),
 \label{eq:mmiss:fitfdata}
\end{eqnarray}
where $N$ represents the yield of each component and first (second) term in the equation describes contribution of signal (sum of five background components).
Histogram PDF of signal, ${\cal H}_{\rm sig}^{\xfrag}(\mmiss(\dtagkx\gamma))$, is additionally
numerically convolved with a Gaussian function, ${\cal G}(\sigma_{\rm cal})$ -- centered at zero and with width $\sigma_{\rm cal}$, which takes
into account possible differences between $\mmiss(\dtagkx\gamma)$ resolutions obtained on real data and simulated samples.

The real data resolution in $\mmiss(\dtagkx\gamma)$ for signal candidates is calibrated using the mass difference 
between $\dsst$ and $\ds$, $\Delta M = M_{\dsst} - M_{\ds}$, for exclusively reconstructed $\dssttodsg$ decays, 
where $\ds$ decays to $\phi\pi$ and $\phi\to K^+K^-$. In the exclusive reconstruction of $\dsst$ mesons the same requirements 
are used for the signal photon candidate as in the inclusive reconstruction. The dominant contribution to the resolution of $\Delta M$ as well as 
of the $\mmiss(\dtagkx\gamma)$ is the signal photon energy resolution. In former case the smearing of the $\ds$ momentum cancels almost 
completely in the mass difference while in the latter case the impact of experimental smearing of $p_{\rm miss}(\dtagkx)$ on 
$\mmiss(\dtagkx\gamma)$ is minimized by performing the mass constrained vertex of $\dtagkx$ candidates to the nominal $\dsst$ mass.
According to simulated sample the resolutions of $\Delta M$ and $\mmiss(\dtagkx\gamma)$ are same to within few percent which justifies 
calibration of $\mmiss(\dtagkx\gamma)$ resolution by comparing $\Delta M$ resolutions of exclusively reconstructed $\dssttodsg$ decays
obtained on real data and MC.

In order to calibrate $\mmiss(\dtagkx\gamma)$ resolution a fit to the real data $\Delta M$ distribution is performed which is described 
as the sum of signal and background components:
\begin{equation}
 {\cal F}^{\rm excl}(\Delta M)=N_{\rm sig}{\cal H}_{\rm sig}(\Delta M)\otimes {\cal G}(\sigma_{\rm cal}) + N_{\rm bkg}(1+c_1\Delta M + c_2\Delta M^2).
 \label{eq:deltam:fitf}
\end{equation}
The signal contribution, ${\cal H}_{\rm sig}(\Delta M)\otimes {\cal G}(\sigma_{\rm cal})$, is parametrized using the histogram PDF obtained on
simulated sample, ${\cal H}_{\rm sig}(\Delta M)$, and is numerically convolved with a Gaussian function, ${\cal G}(\sigma_{\rm cal})$.
The background is parametrized as the 2nd order polynomial. Best agreement between real data and simulated $\Delta M$ distributions are
obtained when $\sigma_{\rm cal}=2.0\pm0.2$~MeV.

\begin{figure}[t!]
 \centering
 \includegraphics[width=0.45\textwidth]{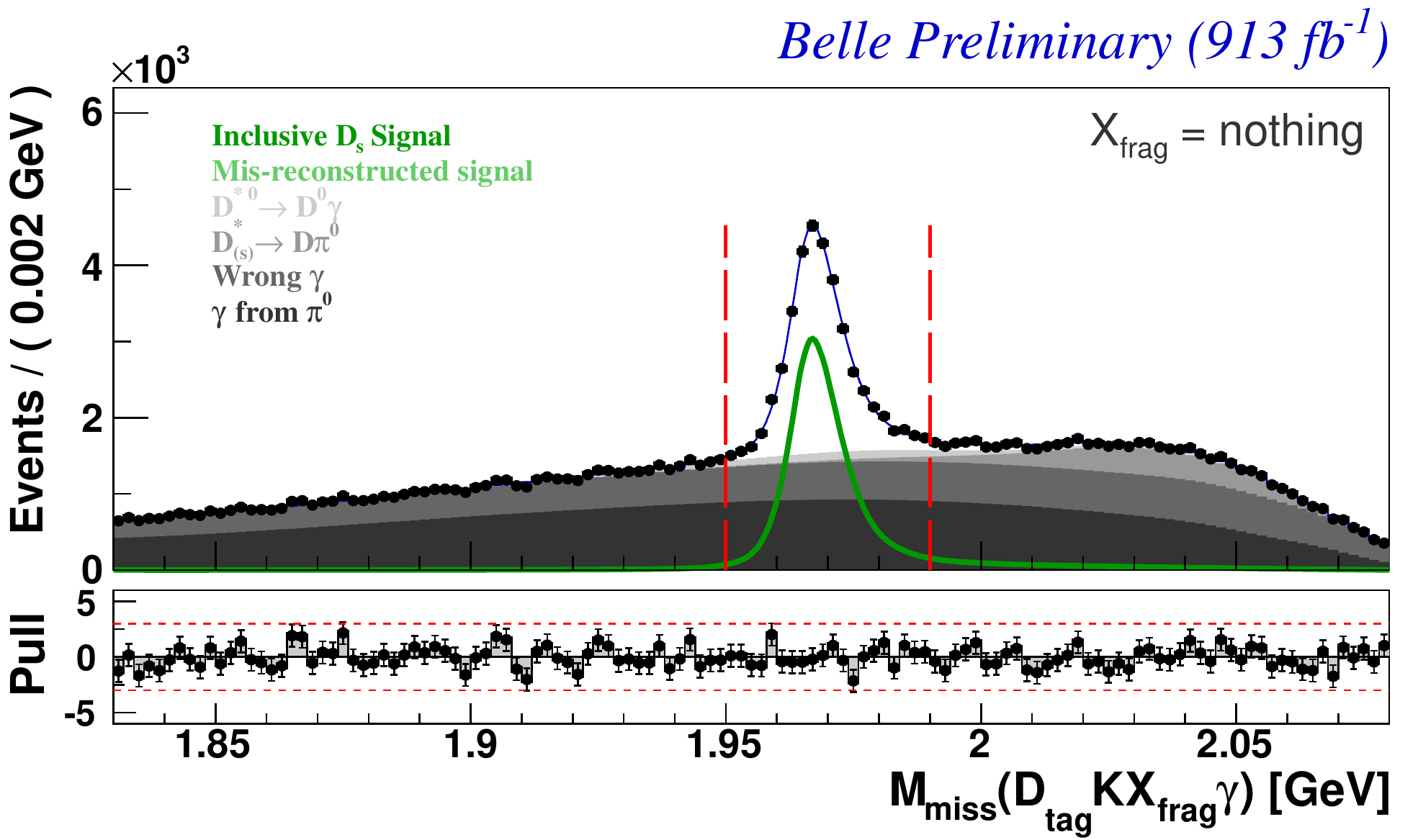}\\
% (a) $\xdmA$\\
 \includegraphics[width=0.45\textwidth]{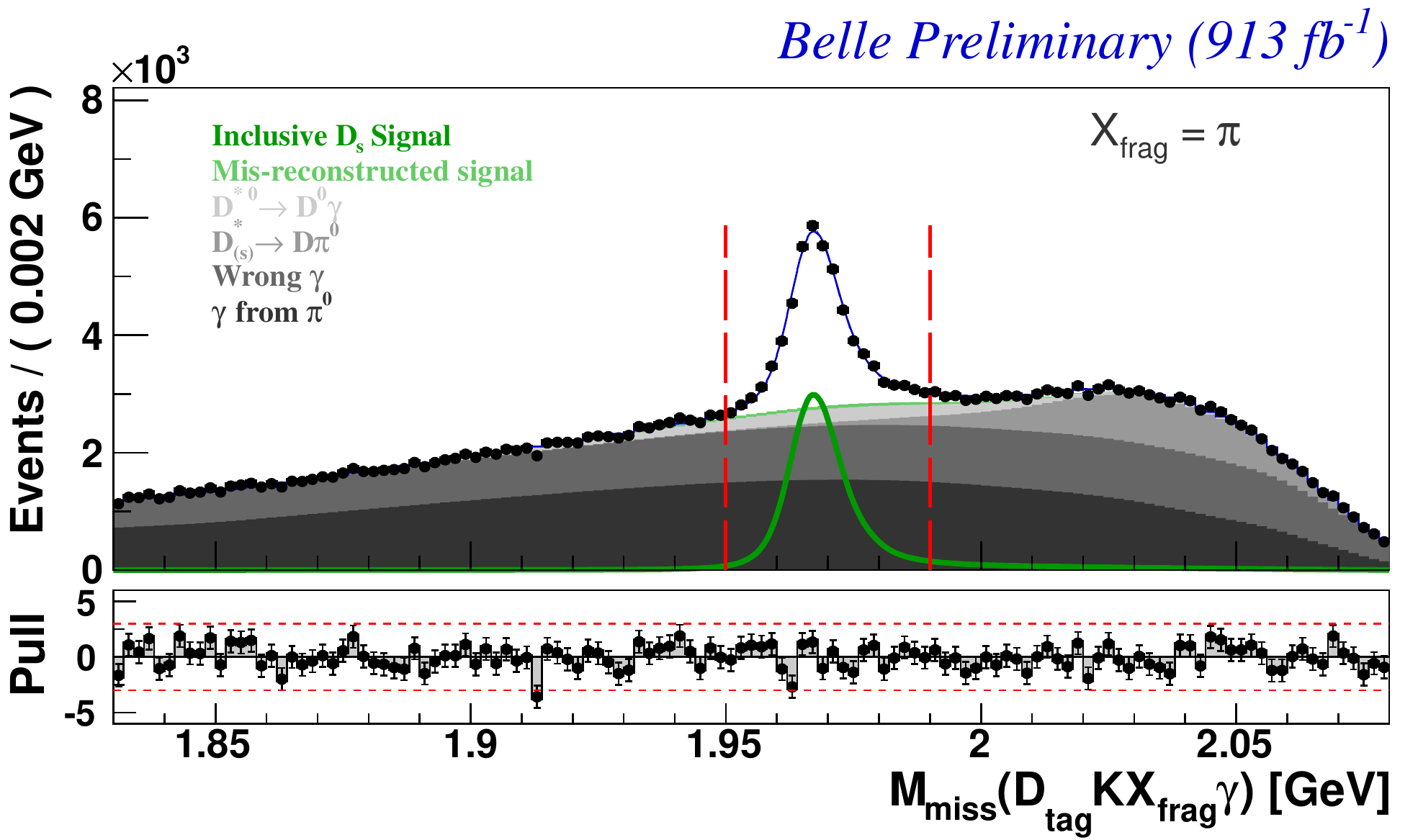}
 \includegraphics[width=0.45\textwidth]{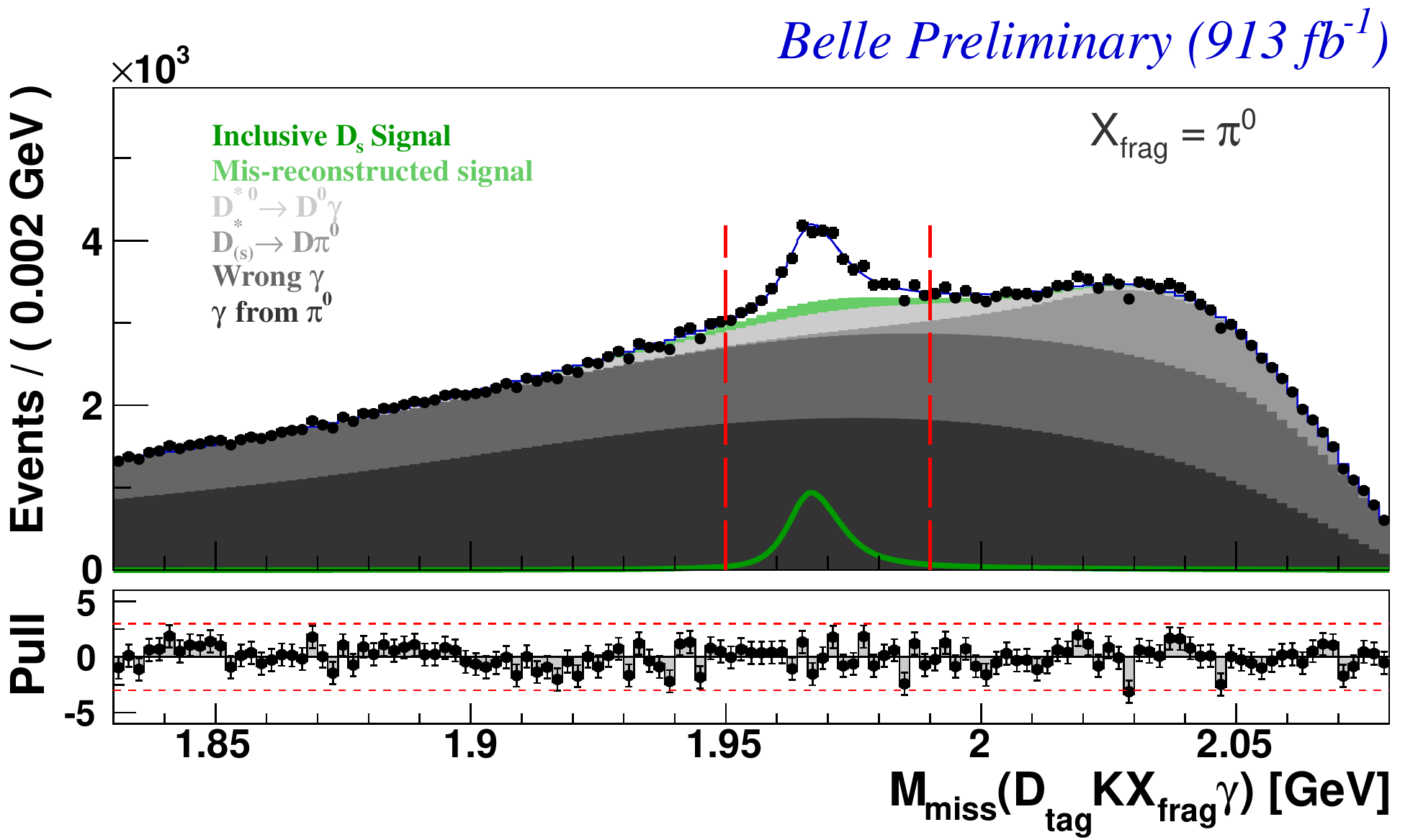}\\
% (b) $\xdmB$\hspace{0.3\textwidth}(c) $\xdmC$\\
 \includegraphics[width=0.45\textwidth]{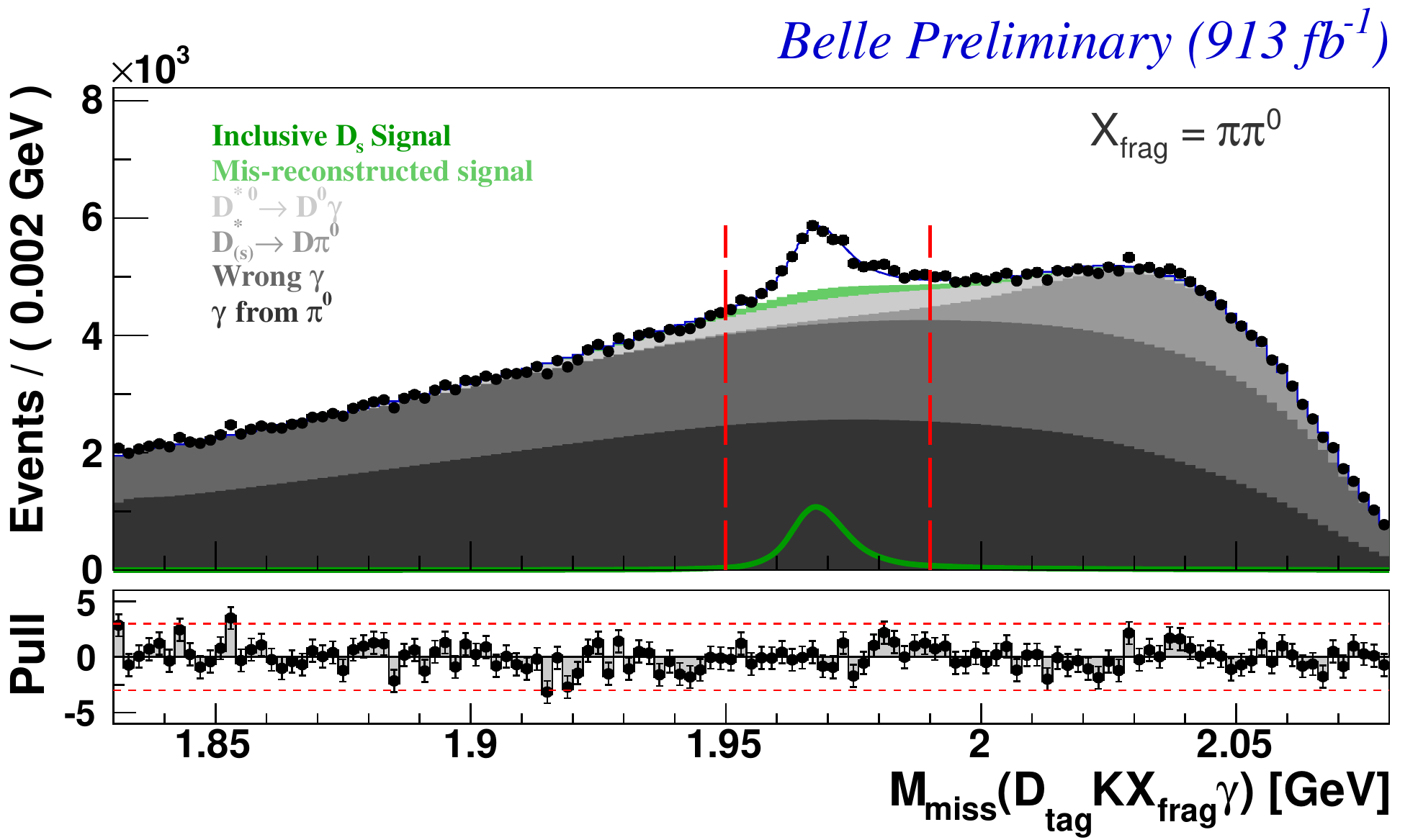}
 \includegraphics[width=0.45\textwidth]{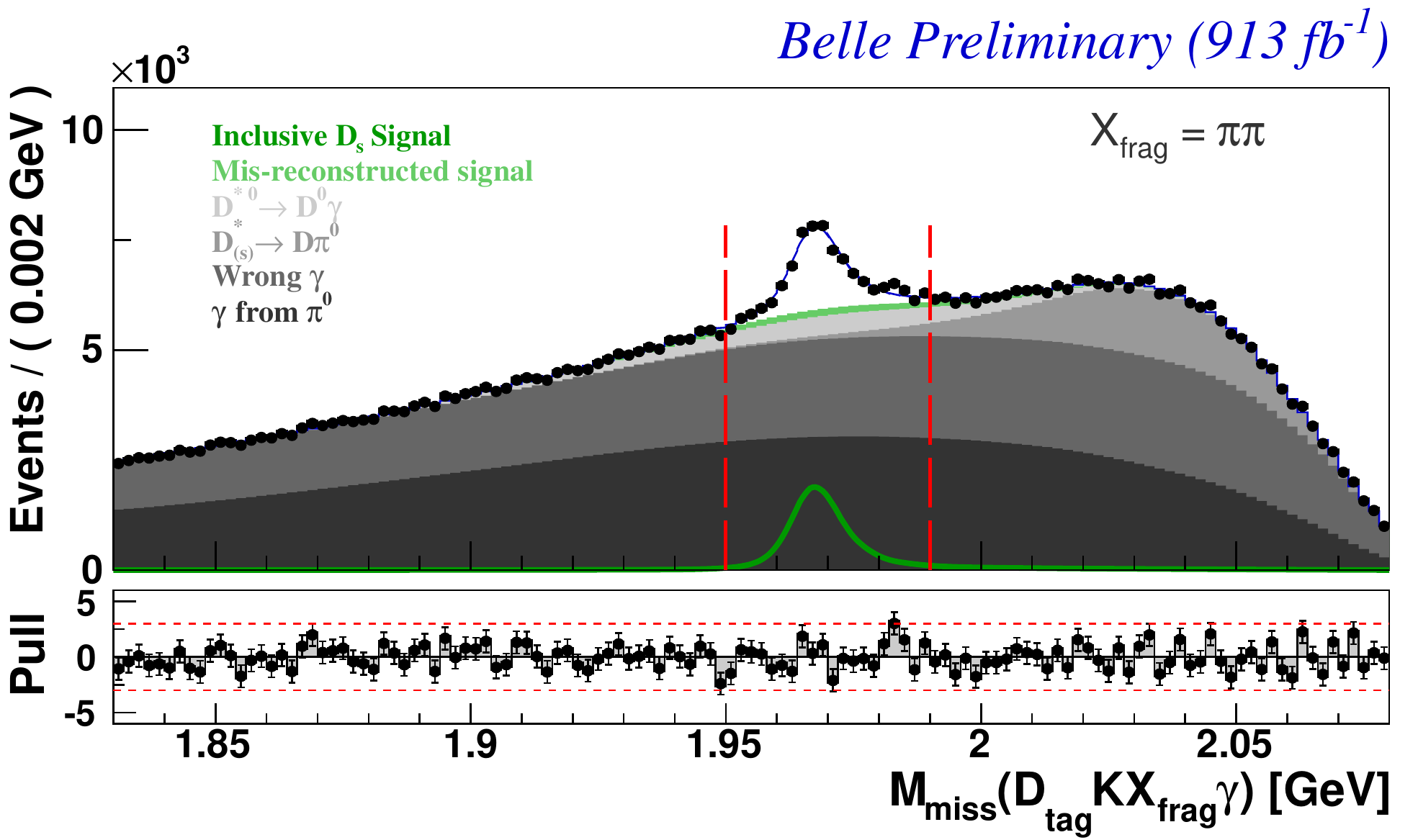}\\
% (d) $\xdmD$\hspace{0.3\textwidth}(e) $\xdmE$\\
 \includegraphics[width=0.45\textwidth]{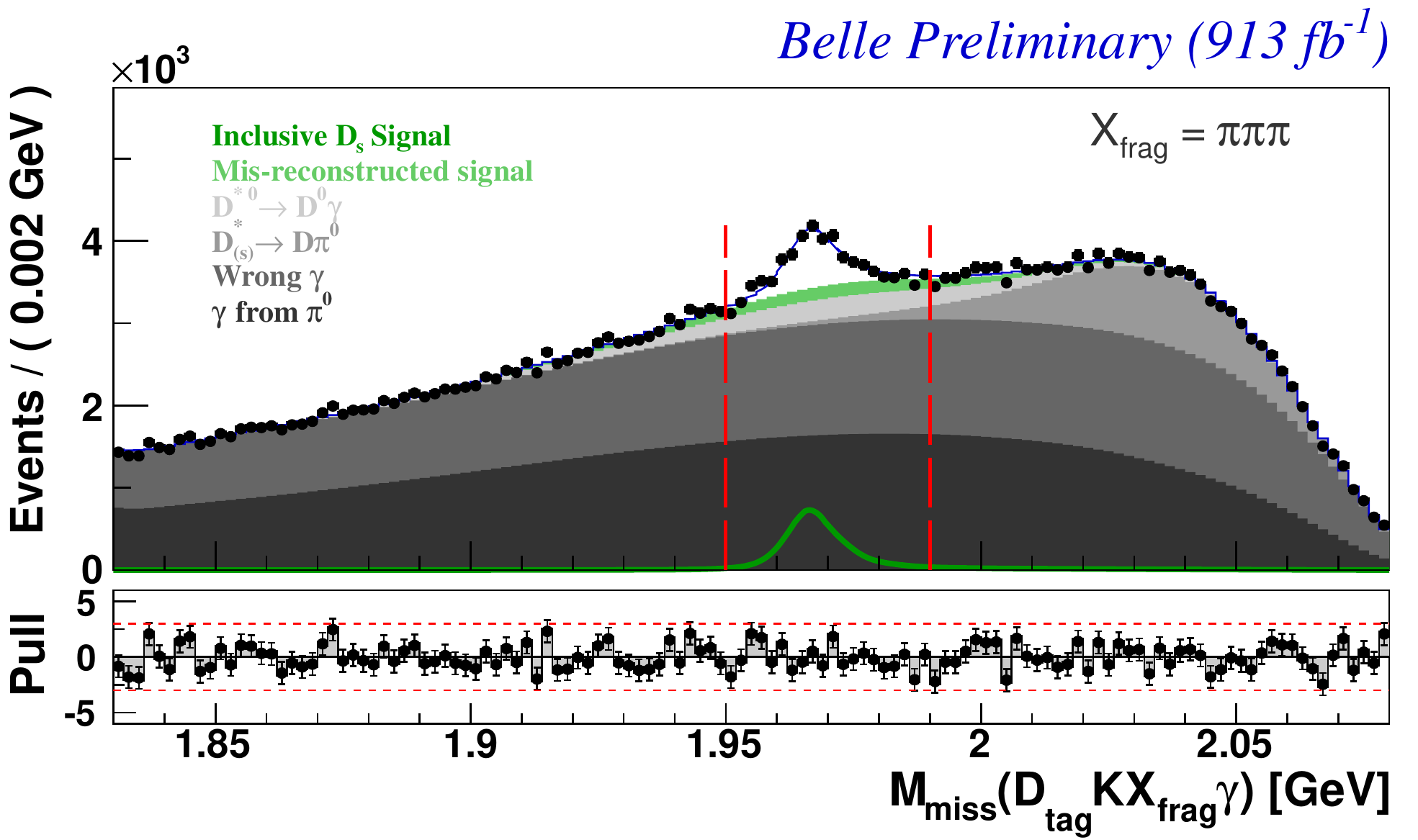}
 \includegraphics[width=0.45\textwidth]{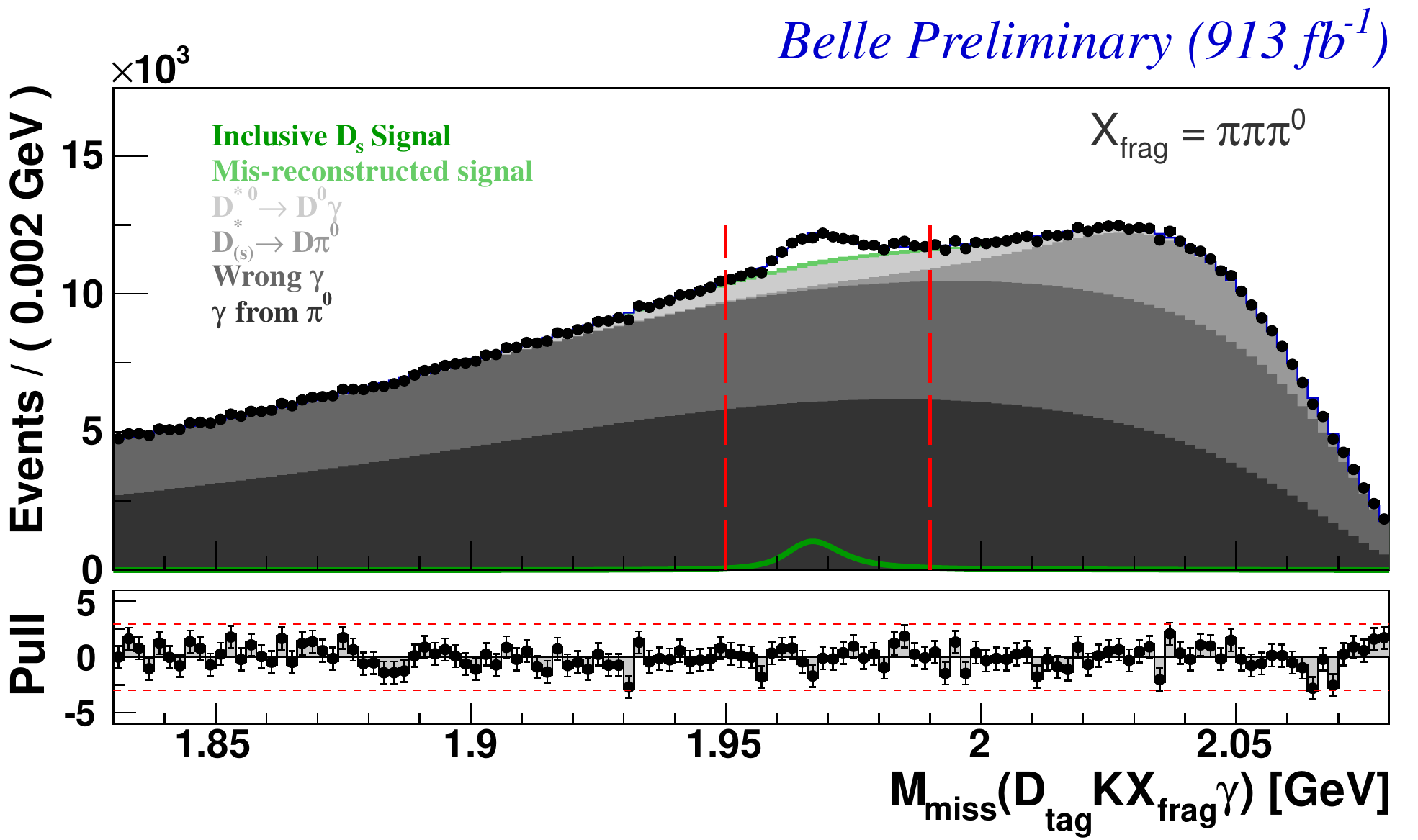}\\
% (f) $\xdmF$\hspace{0.3\textwidth}(g) $\xdmG$\\
 \caption{The $\mmiss(\dtagkx\gamma)$ distributions for all seven $\xfrag$ modes with fit results superimposed (solid blue line). Signal 
 contribution is shown with the solid green line, while the full histograms (in different shades of gray) show the contributions of different
 types of background described in more details in text. The two dashed vertical lines indicate the signal region.}
 \label{fig:dsincl:datafitres}
\end{figure}
The $\mmiss(\dtagkx\gamma)$ distributions for each $\xfrag$ mode of real data inclusive $\ds$ sample with fit results superimposed 
(see Eq.~\ref{eq:mmiss:fitfdata}) are shown in Fig. \ref{fig:dsincl:datafitres}. Total inclusive $\ds$ yield on real data sample 
corresponding to 913 \fb~is found to be $94400\pm 1300$, where uncertainty is statistical only. The PDFs describes well the observed 
data distributions -- the normalized $\chi^2$ values of the fits are between 1.06 and 1.32. Fit residuals exhibit no structures. 
To estimate the systematic uncertainty on the inclusive $\ds$ yield the fits are repeated by taking $\sigma_{\rm cal}=1.8$ or 2.2 MeV. 

The inclusive $\ds$ yield, including systematics uncertainties, is found to be
\begin{equation}
 N^{\rm inc}_{\ds} = 94400\pm1300(\rm stat.)\pm1400(\rm syst.).
\label{eq:inclds:yield}
\end{equation}
We keep only inclusive $\ds$ candidates within the signal window of the $\mmiss(\dtagkx\gamma)$, 
defined as $1.95 < \mmiss(\dtagkx\gamma) < 1.99$~GeV, in rest of the analysis, except in case of exclusive reconstruction of 
$\ds\to K^-K^+\pi^+$ decays.

\section{\boldmath Reconstruction of $\ds\to f$ decays and extraction of absolute $\br(\ds\to f)$}
\label{sec:brandyields}
After reconstructing the inclusive sample of $\ds$ mesons we proceed with the reconstruction of $\ds$ mesons decaying to:
\begin{itemize}
 \item $\ds^+\to K^-K^+\pi^+$,
 \item $\ds^+\to \overline{K}{}^0K^+$,
 \item $\ds^+\to \eta \pi^+$,
 \item $\dsmunu$,
 \item $\ds^+\to \tau^+\nu_{\tau};~\tau^+\to e^+\nu_{e}\overline{\nu}{}_{\tau}$, $\mu^+\nu_{\mu}\overline{\nu}{}_{\tau}$, $\pi^+\overline{\nu}{}_{\tau}$.
\end{itemize}
In the following subsections we briefly describe the reconstruction procedure and signal yield extraction for all five studied decay modes,
which then enters the calculation of absolute $\br(\ds\to f)$ as:
\begin{equation}
 \br(\ds\to f) = \frac{N(\ds\to f)}{N^{\rm inc}_{\ds}\cdot f_{\rm bias}\cdot \varepsilon(\ds\to f|{\rm incl.}~\ds)}.
 \label{eq:abs_br_ideal}
\end{equation}
Above, $N^{\rm inc}_{\ds}$ is the number of inclusively reconstructed $\ds$ mesons, $N(\ds\to f)$ is 
the number of reconstructed $\ds\to f$ decays within the inclusive $\ds$ sample, and $\varepsilon(\ds\to f|{\rm inc.}~\ds)$ is the efficiency of
reconstructing $\ds\to f$ decay within the inclusive $\ds$ sample. We observed on simulated samples that the efficiency of 
inclusive reconstruction of $\ds$ mesons depends on the $\ds$ decay mode and therefore the inclusively reconstructed $\ds$ sample 
does not represent truly inclusive sample of $\ds$ mesons. The efficiency drops with increasing 
multiplicity of final state particles, e.g. $N_{\rm ch} + N_{\pi^0}$ (where $N_{\rm ch}$ represents the number of charged particles and
$N_{\pi^0}$ the number of neutral pions), produced in $\ds$ decays. In order to take this effect into account a ratio
of efficiencies to reconstruct $\ds$ meson inclusively if it decayed to final state $f$, 
$\varepsilon^{\rm inc}_{\ds\to f}$, and the average efficiency of inclusive $\ds$ reconstruction, 
$\overline{\varepsilon}{}_{\ds}^{\rm inc}=\sum_i {\cal B}(\ds\to i)\varepsilon^{\rm inc}_{\ds\to i}$ is included into the
denominator of Eq. \ref{eq:abs_br_ideal}: $\fbias = {\varepsilon^{\rm inc}_{\ds\to f}}/{\overline{\varepsilon}{}_{\ds}^{\rm inc}}$. Ratio 
$\fbias$ is taken from the simulated sample including all known $\ds$ decay modes. 

Possible differences between simulated and real data samples in terms of $\ds$ decay modes and their $\br$s used in our simulation 
are estimated by studying the distributions of number of particles, $N_{\rm ch} + N_{\pi^0}$, produced in $\ds$ decays. This distribution
on real data is obtained in the following way: first for each inclusive $\ds$ candidate the number of remaining charged tracks, 
$N^{\rm reco}_{\rm ch}$, and $\pi^0$ candidates, $N_{\pi^0}^{\rm reco}$, not associated to $\dtagkx\gamma$ candidate are counted; second
fits to the $\mmiss(\dtagkx\gamma)$ distributions are performed in bins of $N^{\rm reco}_{\rm ch} + N^{\rm reco}_{\pi^0}$ and the 
inclusive $\ds$ signal yield is recorded. The distribution of $N^{\rm reco}_{\rm ch} + N^{\rm reco}_{\pi^0}$ is roughly proportional to the
true distribution of $N_{\rm ch}+N_{\pi^0}$, but with a considerable amount of convolution\footnote{E.g. $\ds$ daughter particle might 
not be reconstructed or a fake charged track or $\pi^0$ candidate is counted.}. The unfolded $N_{\rm ch} + N_{\pi^0}$ distributions are 
obtained from $N^{\rm reco}_{\rm ch} + N^{\rm reco}_{\pi^0}$ distribution using the singular value decomposition algorithm. From 
the distributions of $N_{\rm ch} + N_{\pi^0}$ obtained on real data and simulated samples and the dependence of inclusive
$\ds$ reconstruction efficiency obtained on simulated sample we estimate the the ratio between 
real data and simulated average inclusive $\ds$ reconstruction efficiencies to be:
${\overline{\varepsilon}{}_{\ds}^{\rm inc}|_{\rm DATA}}/{\overline{\varepsilon}{}_{\ds}^{\rm inc}|_{\rm MC}}=0.9768\pm0.0134$.
The ratio is consistent with unity within the uncertainty. Nevertheless the inverse of the $\fbias$ factor is corrected by the 
above ratio and its error is taken as a source of systematic uncertainty of measured absolute branching fractions.

\subsection{\boldmath $\ds^+\to K^-K^+\pi^+$}

The reconstruction of $\ds^+\to K^-K^+\pi^+$ decays is performed by requiring exactly three charged tracks in the rest of the event 
with a net charge equal to the charge of inclusively reconstructed $\ds$ candidate. The track with opposite charge to the inclusive $\ds^+$ 
candidate is selected to be $K^-$ candidate while the two same-sign tracks are identified as $K^+$ or $\pi^+$ based on their 
likelihood ratios, ${\cal L}_{K,\pi}$.

The exclusively reconstructed  $\ds^+\to K^-K^+\pi^+$ candidates are identified as a peak at the nominal mass of the $\ds^{\ast +}$ in the
invariant mass distribution of $K^+K^-\pi^+\gamma$ combination, $M(K^+K^-\pi^+\gamma)$. Here the $\gamma$ stands for the signal
photon candidate used to reconstruct $\dtagkx\gamma$ candidate. The $M(K^+K^-\pi^+\gamma)$ is chosen over $M(K^+K^-\pi^+)$, because 
both sides (inclusive and exclusive) have to be correctly reconstructed to produce peaks in $M(K^+K^-\pi^+\gamma)$ and 
$\mmiss(\dtagkx\gamma)$. Correctly reconstructed $\ds^+\to K^-K^+\pi^+$ events will namely peak in $M(K^+K^-\pi^+)$ even if 
the inclusive reconstruction of $\ds$ candidates failed, e.g. the photon candidate which does not originate from $\dssttodsg$ 
decay is picked up. Since $M(K^-K^+\pi^+\gamma)$ and  $\mmiss(\dtagkx\gamma)$ are correlated due to their common input -- $\gamma$ 4-momentum -- 
no additional selection is applied to the $\mmiss(\dtagkx\gamma)$. 

The signal yield of exclusively reconstructed $\ds\to KK\pi$ decays is extracted by fitting the $M(K^+K^-\pi^+\gamma)$ distribution. The 
components of the fit are divided into three categories: correctly reconstructed $\dsst\to\ds\gamma\to KK\pi\gamma$ decays (signal);
mis-reconstructed $\dsst\to\ds\pi^0\to K^+K^-\pi^+\gamma\gamma$ decays, where one of the photons from the $\pi^0$ decay is not reconstructed
($\dsst\to\ds\pi^0\to KK\pi\gamma\gamma$); and random combinations of charged tracks and photons (combinatorial background). The 
$M(K^+K^-\pi^+\gamma)$ distribution is parameterized as:
\begin{eqnarray}
 {\cal F}(M(K^+K^-\pi^+\gamma)) & = & N_{\rm sig}\cdot {\cal H_{\rm sig}}(M(K^+K^-\pi^+\gamma))\otimes {\cal G}(\sigma_{\rm cal}^{\rm excl})\nonumber\\
 & & + N_{\dsst\pi^0}\cdot {\cal H}_{\dsst\pi^0}(M(K^+K^-\pi^+\gamma))\nonumber\\
 & & + N_{\rm comb}\cdot (1+c_1\cdot M(K^+K^-\pi^+\gamma) +\nonumber\\&& c_2\cdot M(K^+K^-\pi^+\gamma)^2+c_3\cdot M(K^+K^-\pi^+\gamma)^3 ),
\end{eqnarray}
where the signal and $\dsst\to\ds\pi^0$ background are parametrized using the non-parametric histogram PDF obtained on a 
simulated sample (${\cal H}$), and the combinatorial background is parametrized as a third order polynomial. 
The signal distribution is convolved with a Gaussian function, ${\cal G}(\sigma_{\rm cal}^{\rm excl})$, in order to take into account
the differences between the resolutions of $M(K^+K^-\pi^+\gamma)$ on real data and simulated samples. The $\sigma_{\rm cal}^{\rm excl}$ is 
estimated using the same procedure as described in section \ref{sec:inclds:yield:extraction}. The only difference is that the resolution 
on $M(KK\pi\gamma)$ is calibrated instead of the $\dsst$ and $\ds$ mass difference. We determine $\sigma_{\rm cal}^{\rm excl} = 3.2\pm0.2$ MeV. 
Free parameters of the fit are the normalization parameters, $N_i$, and combinatorial background shape parameters, $c_i$.

\begin{figure}[t]
  \centering
  \includegraphics[width=1\textwidth]{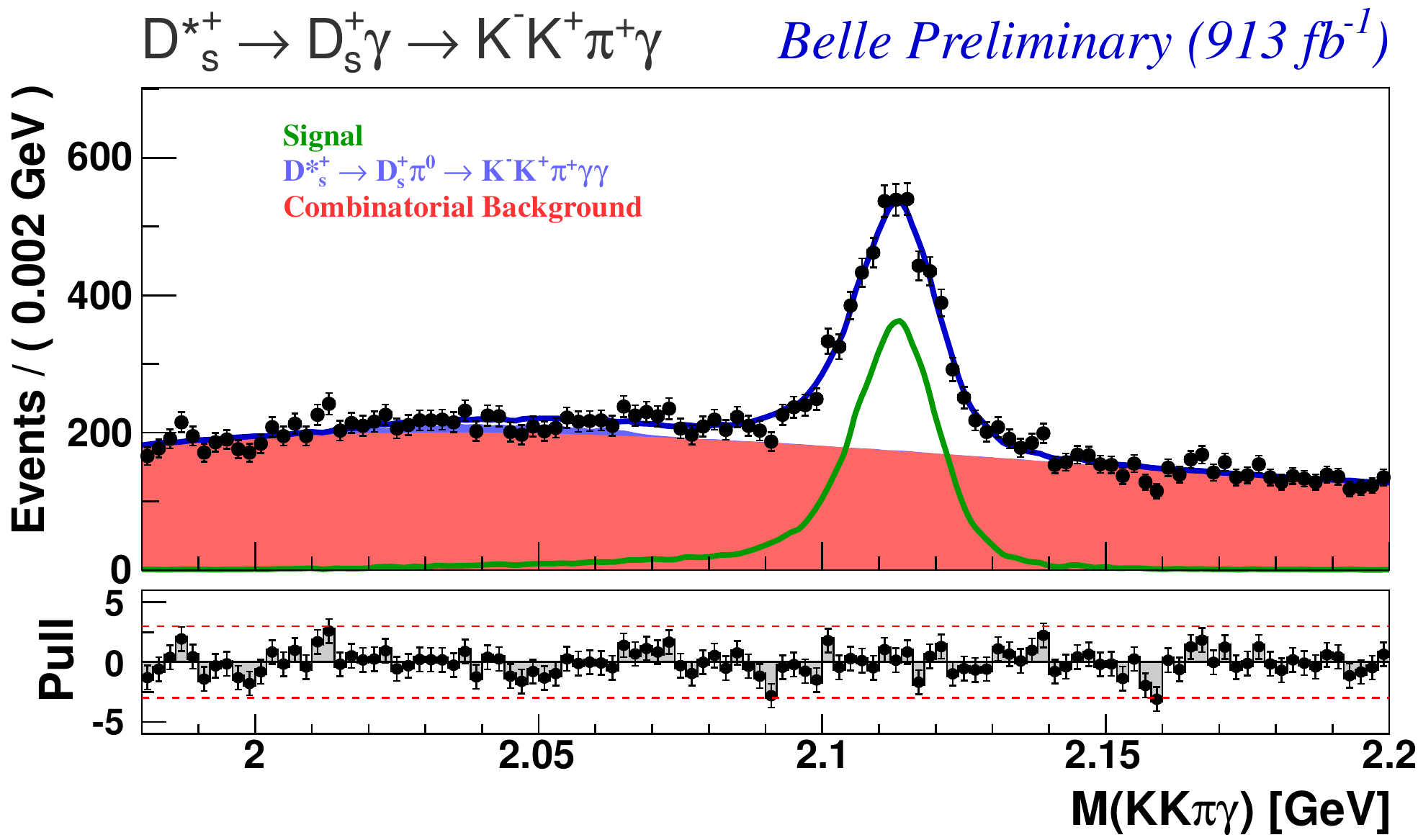} 
  \caption{The $M(K^+K^-\pi^+\gamma)$ distribution of exclusively reconstructed $\ds\to KK\pi$ decays within the inclusive $\ds$ sample 
  with fit results superimposed (solid blue line). Solid green line shows the signal contribution, while the full histograms show the contribution of
  combinatorial background (in red) and $\ds\to KK\pi$ decays originating from $\dsst\to\ds\pi^0$ decay (in blue).}
  \label{figs:kkpi:daresults}
\end{figure}

The $M(K^+K^-\pi^+\gamma)$ distribution of exclusively reconstructed $\ds\to KK\pi$ decays within the inclusive $\ds$ sample
obtained on real data sample is shown in Fig. \ref{figs:kkpi:daresults} with fit results superimposed. 
The number of correctly reconstructed $\ds\to KK\pi$ decays is found to be
\begin{equation}
 N(\ds\to KK\pi)=4094\pm123,
 \label{eq:dskkpi:yieldDA}
\end{equation}
where the error is statistical only.

\subsection{\boldmath $\ds^+\to \overline{K}{}^0K^+$}

The $\ds^+\to \overline{K}{}^0K^+$ decays are reconstructed partially %within the signal window of the inclusive $\ds$ sample 
%($1.95<\mmiss(\dtagkx\gamma)<1.99$ GeV). 
by requiring only one additional charged track to be present in rest of the event. The charged track is required to be consistent
with kaon hypothesis and has charge equal to that of the inclusively reconstructed $\ds$ candidate. The neutral kaon is not
reconstructed at all and it is identified as a peak at the nominal mass of neutral kaon squared in the missing mass squared distribution:
\[
 \mmiss^2(\dtagkx\gamma K) = p_{\rm miss}^2(\dtagkx\gamma K),
\]
where the missing 4-momentum is given by
\[
 p_{\rm miss}(\dtagkx\gamma K)= p_{e^+} + p_{e^-} - p_{\dtag} - p_{K} - p_{\xfrag} - p_{\gamma} - p_{K}.
\]
An explicit reconstruction of $\overline{K}{}^0$ meson (via the experimentally accessible $K^0_S\to\pi^+\pi^-$ decays) 
would lead to a significant signal loss ($2/3$ of the signal would be lost, without even taking into account the $K^0_S$ 
reconstruction efficiency). The signal peak of $\ds^+\to \overline{K}{}^0K^+$ in $\mmiss^2(\dtagkx\gamma K)$ distribution is in addition 
used to study the differences between $\mmiss^2$ resolutions in simulated and real data samples, which is important in the extraction of 
yield of $\dsmunu$ decays.
Since the flavor of the neutral kaon is not determined, the doubly Cabibbo suppressed decays, $\ds^+\to K^0K^+$, also contribute to the peak in 
$\mmiss^2(\dtagkx\gamma K)$. Their relative contribution can naively be estimated to be equal to $\tan^4\theta_C\approx 0.29\%$ ($\theta_C$
being the Cabibbo mixing angle), which is order of magnitude below the expected statistical uncertainty and can thus be safely neglected.

The signal yield of partially reconstructed $\ds^+\to \overline{K}{}^0K^+$ decays is extracted by performing a fit to the 
$\mmiss^2(\dtagkx\gamma K)$ distribution. The candidates are divided into six categories: 
$K^+$ candidate originates from the $\ds^+\to \overline{K}{}^0K^+$ decay and the inclusive $\ds$ candidate is correctly reconstructed (signal);
$K^+$ candidate originates from the $\ds^+\to \eta K^+$ decay and the inclusive $\ds$ candidate is correctly reconstructed 
($\ds\to \eta K$ decays);
$K^+$ candidate originates from the $\ds^+\to \pi^0 K^+$ decay and the inclusive $\ds$ candidate is correctly reconstructed ($\ds\to \pi^0 K$ decays);
$K^+$ candidate is mis-reconstructed pion originating from the $\ds^+\to \eta \pi^+$ decay 
 and the inclusive $\ds$ candidate is correctly reconstructed ($\ds^+\to \eta \pi^+$); $K^+$ candidate originates from the $\ds\to K^{\ast +} \overline{K}{}^0\to K^+\pi^0 \overline{K}{}^0$ decay 
 and the inclusive $\ds$ candidate is correctly reconstructed ($\ds\to K^{\ast +} \overline{K}{}^0$ decays); and all other candidates (combinatorial
 background). The $\mmiss^2(\dtagkx\gamma K)$ distribution is parameterized as:
\begin{eqnarray}
 {\cal F}(\mmiss^2) & = & N_{\rm sig}\cdot \sum_{i=1}^{3}{\cal G}(\mmiss^2,m^2_{K^0},s_i\sigma^{MC}_i)\nonumber\\
 & & + N_{\eta K}\sum_{i=1}^{3}{\cal G}(\mmiss^2,m^2_{\eta},\sigma_i^{MC}) + N_{\pi^0 K}\sum_{i=1}^{3}{\cal G}(\mmiss^2,m^2_{\pi^0},\sigma_i^{MC})\nonumber\\
 & & + N_{\eta \pi}[{\cal G}(\mmiss^2, m_0^{\eta\pi},\sigma_G)+{\cal BG}(\mmiss^2,m_0^{\eta\pi},\sigma_{BG}^{\rm left},\sigma_{BG}^{\rm right})]\nonumber\\
 & & + N_{K^{\ast}K^0}[{\cal G}(\mmiss^2, m_0^{K^{\ast}K^0},\sigma_G)+{\cal BG}(\mmiss^2,m_0^{K^{\ast}K^0},\sigma_{BG}^{\rm left},\sigma_{BG}^{\rm right})]\nonumber\\
 & & + N_{\rm comb}\cdot (1+c_1\mmiss^2 + c_2\mmiss^2+c_3\mmiss^3 +c_4\mmiss^4  ),
\end{eqnarray}
where the signal, and the two peaking backgrounds ($\ds\to \eta K$ and $\ds\to \pi^0 K$) are parametrized using the sum of 3 Gaussian functions. 
All the parameters of the latter peaking backgrounds are fixed to values determined
on the simulated sample. In case of the signal all the shape parameters are fixed to the values determined on simulated sample, 
except the mean ($m^2_{K^0}$) and the common resolution scaling factor of the core and the second Gaussian $(s_1=s_2=s)$. 
The width of the third Gaussian (describing the outliers) is fixed. The $\eta\pi$ and $K^{\ast}K^0$ candidates are parametrized 
with a sum of Gaussian and a Bifurcated Gaussian function, where all parameters are fixed to the values determined on simulated sample. 
The combinatorial background is parametrized with a polynomial of the $4^{th}$ order, where the coefficients $c_i$ are determined with 
the fit to the $\mmiss^2(\dtagkx\gamma K)$ distribution for candidates in the  $\mmiss(\dtagkx\gamma)<1.95$ GeV sideband region. Yields of 
all event categories are free parameters of the fit, except for the $\ds\to \eta K$ and $\ds\to \eta \pi$, which are constrained to 
expected values based on their known $\br$s and MC determined efficiencies.

\begin{figure}[t]
 \centering
 \includegraphics[width=1\textwidth]{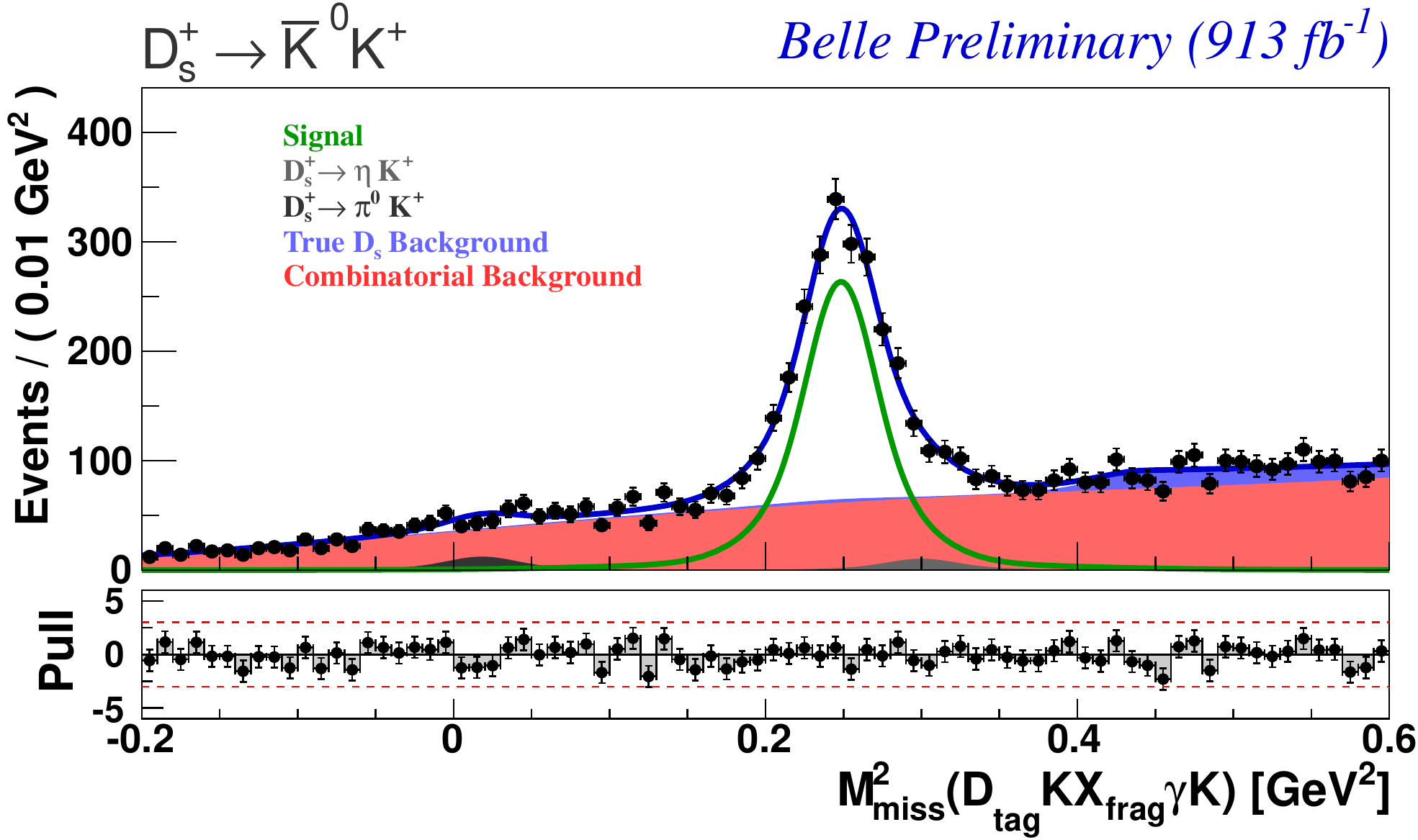} 
 \caption{The $\mmiss^2(\dtagkx\gamma K)$  distribution of partially reconstructed $\ds^+\to \overline{K}{}^0K^+$ decays within 
the inclusive $\ds$ sample with fit results superimposed (solid blue line). Solid green line shows the signal contribution, 
while the full histograms show the contribution of combinatorial background (in red), $\ds^+\to \pi^0 K^+$ (dark gray), 
$\ds^+\to \eta K^+$ (light gray) or other $\ds^+$ decays (blue).}
\label{figs:kk0:daresults}
\end{figure}

The  $\mmiss^2(\dtagkx\gamma K)$ distribution of partially reconstructed  $\ds^+\to \overline{K}{}^0K^+$ decays 
within the inclusive $\ds$ sample obtained on real data sample is shown in Fig. \ref{figs:kk0:daresults} with fit results superimposed. 
The number of correctly reconstructed $\ds^+\to \overline{K}{}^0K^+$ decays is found to be
\begin{equation}
 N(\ds\to K^0 K)=1943\pm82,
 \label{eq:dskk0:yieldDA}
\end{equation}
where the error is statistical only.

\subsection{\boldmath $\ds^+\to \eta \pi^+$}

As in case of $\ds^+\to \overline{K}{}^0K^+$ decays we perform partial reconstruction of $\ds^+\to \eta \pi^+$ decays as well. We require 
only one charged track consistent with the pion hypothesis to be present in rest of the event. We do not perform any explicit reconstruction of 
$\eta$ mesons which are identified as a peak at the nominal mass squared of $\eta$ in the missing mass squared distribution:
\[
 \mmiss^2(\dtagkx\gamma \pi) = p_{\rm miss}^2(\dtagkx\gamma \pi),
\]
where the missing 4-momentum is given by
\[
 p_{\rm miss}(\dtagkx\gamma \pi)= p_{e^+} + p_{e^-} - p_{\dtag} - p_{K} - p_{\xfrag} - p_{\gamma} - p_{\pi}.
\]
An explicit reconstruction of $\eta$ meson would lead to a significant signal loss. 

In this sample of inclusive $\ds$ candidate plus additional charged pion there is a significant contribution from $\ds\to\tau\nu$ decays, when
$\tau$ lepton decays hadronically to charged pion and neutrino. These events are suppressed by requiring that the extra neutral energy in the ECL 
($E_{\rm ECL}$) to be larger than 1.0 GeV, where the $E_{\rm ECL}$ represents a sum over all energy deposits in the ECL which are not associated
to the tracks or neutrals used in inclusive reconstruction of $\ds$ candidates nor the charged pion candidate \cite{Ikado:2006un}. The 
$\ds\to\tau\nu\to\pi\nu\nu$ decays namely peak in $E_{\rm ECL}$ at 0, while $\ds\to\eta\pi$ decays deposit significant 
amount of energy in the ECL via the $\eta$ decay products (see section \ref{sec:dstaunu} for more details).

The signal yield of partially reconstructed $\ds^+\to \eta \pi^+$ decays is extracted by performing a fit to the 
$\mmiss^2(\dtagkx\gamma \pi)$ distribution. The candidates are divided into five categories: $\pi^+$ candidate originates 
from the $\ds^+\to \eta \pi^+$ decay and the inclusive $\ds$ candidate is correctly reconstructed (signal); 
$\pi^+$ candidate originates from the $\ds^+\to K^0\pi^+$ decay and the inclusive $\ds$ candidate is correctly reconstructed ($\ds\to K^0 \pi$ decays);
$\pi^+$ candidate is mis-reconstructed kaon originating from the $\ds^+\to \overline{K}{}^0 K^+$ decay and the inclusive $\ds$ candidate is correctly reconstructed 
($\ds\to \overline{K}{}^0 K$ decays);
$\pi^+$ candidate originates from the $\ds\to \rho^0 K^+\to \pi^+\pi^-K^+$ decay 
 and the inclusive $\ds$ candidate is correctly reconstructed ($\ds\to \rho^0 K^+$ decays); and all other candidates (combinatorial background).
The $\mmiss^2(\dtagkx\gamma \pi)$ distribution is parameterized as:
\begin{eqnarray}
 {\cal F}(\mmiss^2) & = & N_{\rm sig}\cdot \sum_{i=1}^{3}{\cal G}(\mmiss^2,m^2_{\eta},s_i\sigma^{MC}_i)+ N_{K^0\pi}\sum_{i=1}^{3}{\cal G}(\mmiss^2,m^2_{K^0},\sigma_i^{MC})\nonumber\\
 & & + N_{K^0 K}[{\cal BG}(\mmiss^2,m_0^{K^0K},\sigma_{BG}^{\rm left},\sigma_{BG}^{\rm right})]+ N_{\rho K}[{\cal BG}(\mmiss^2,m_0^{\rho K},\sigma_{BG}^{\rm left},\sigma_{BG}^{\rm right})]\nonumber\\
 & & + N_{\rm comb}\cdot (1+c_1\mmiss^2 + c_2\mmiss^2+c_3\mmiss^3 +c_4\mmiss^4  ),
\end{eqnarray}
where the signal and the peaking background from $\ds\to K^0 \pi$ decays are parametrized using the sum of three 
Gaussian functions. All the parameters of the latter peaking background are fixed to values determined
on simulated sample. In case of the signal all the shape parameters are fixed, except the mean ($m^2_{\eta}$). 
The common resolution scaling factor of the core and the second Gaussian $(s_1=s_2=s)$ is constrained to the value obtained from the fit to
the $\ds\to K^0K$ sample. The width of the third Gaussian (describing the outliers) is fixed. The $K^0K$ and $\rho K$ candidates are 
parametrized with a Bifurcated Gaussian function, where all parameters are fixed to the values determined on simulated sample. 
The combinatorial background is parametrized with a polynomial of the $4^{th}$ order, where the coefficients $c_i$ are determined 
with the fit to the $\mmiss^2(\dtagkx\gamma \pi)$ distribution for candidates in the $\mmiss(\dtagkx\gamma)<1.95$ GeV sideband region. 
Yields of all event categories are free parameters of the fit, except for the $K^0\pi$ and $K^0K$, which are constrained to expected values 
based on their known $\br$s and efficiencies determined on simulated sample.

\begin{figure}[t]
 \centering
 \includegraphics[width=1\textwidth]{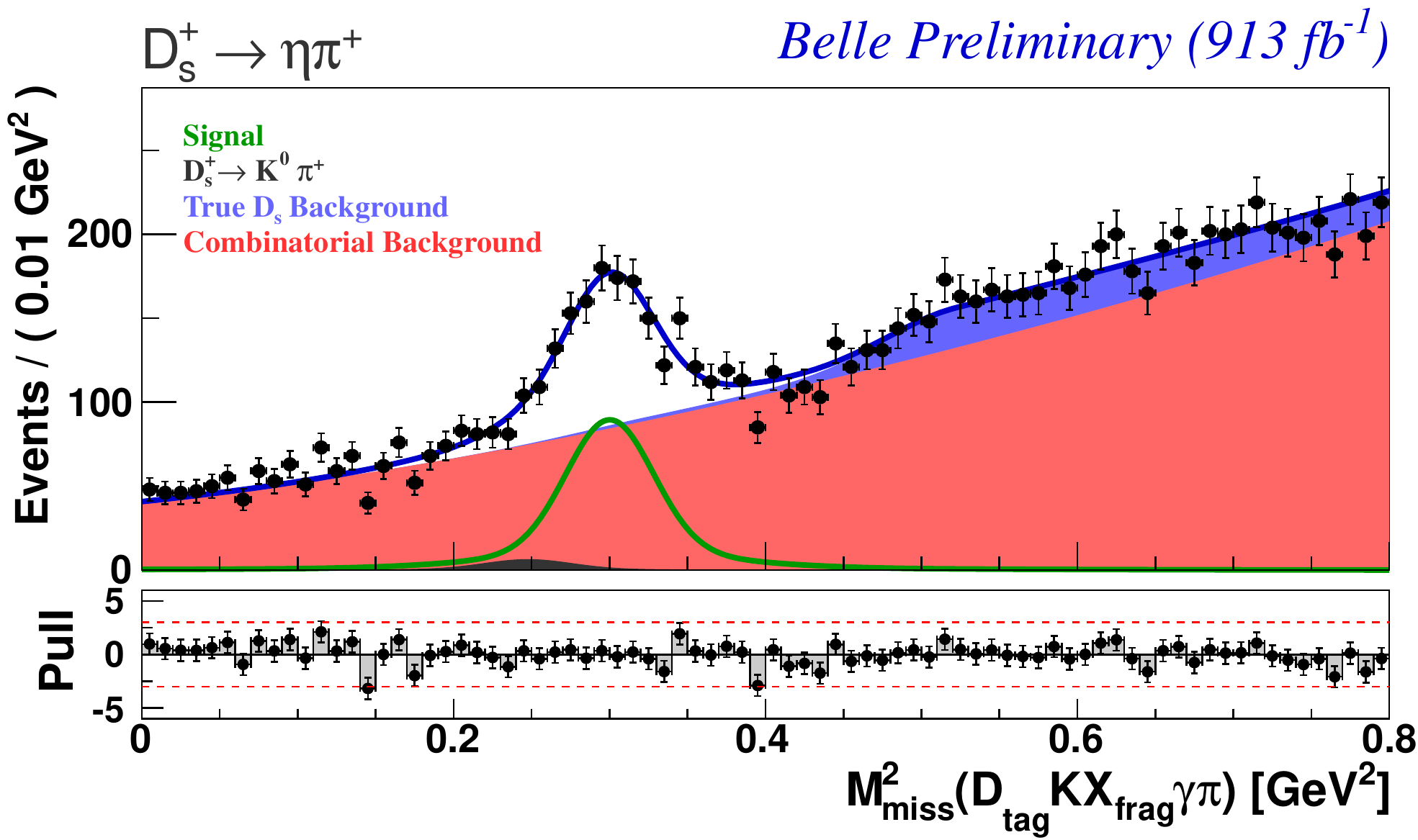} 
 \caption{The $\mmiss^2(\dtagkx\gamma \pi)$  distribution of partially reconstructed $\ds^+\to \eta \pi^+$ decays within 
the inclusive $\ds$ sample with fit results superimposed (solid blue line). Solid green line shows the signal contribution, while the full histograms 
show the contribution of combinatorial background (in red), $\ds^+\to K^0 \pi^+$ (dark gray), or other $\ds^+$ decays (blue).}
 \label{figs:etapi:daresults}
\end{figure}

The  $\mmiss^2(\dtagkx\gamma \pi)$ distribution of exclusively reconstructed  $\ds^+\to \eta\pi^+$ decays 
within the inclusive $\ds$ sample obtained on real data sample is shown in Fig. \ref{figs:etapi:daresults} with fit results superimposed. 
The number of correctly reconstructed $\ds^+\to \eta\pi^+$decays is found to be
\begin{equation}
 N(\ds\to \eta\pi)=773\pm58,
 \label{eq:dsetapi:yieldDA}
\end{equation}
where the error is statistical only.

\subsection{\boldmath $\dsmunu$}
The $\dsmunu$ decays are reconstructed by requiring additional charged track consistent with muon hypothesis to be present in rest of the event. 
Single missing neutrino is then identified as a peak at zero in the missing mass squared distribution:
\[
 \mmiss^2(\dtagkx\gamma \mu) = p_{\rm miss}^2(\dtagkx\gamma \mu),
\]
where the missing 4-momentum is given by
\[
 p_{\rm miss}(\dtagkx\gamma \mu)= p_{e^+} + p_{e^-} - p_{\dtag} - p_{K} - p_{\xfrag} - p_{\gamma} - p_{\mu}.
\]

The signal yield is extracted by performing a fit to the $\mmiss^2(\dtagkx\gamma \mu)$ distribution. The candidates are grouped into five
categories: $\mu^+$ candidate originates from the $\dsmunu$ decay and the inclusive $\ds$ candidate is correctly
 reconstructed (signal); $\mu^+$ candidate is mis-reconstructed kaon originating from the $\ds^+\to \overline{K}{}^0 K^+$ decay and the inclusive $\ds$ candidate is correctly
 reconstructed ($\ds\to K^0 K$ decays);  $\mu^+$ candidate is mis-reconstructed pion originating from the $\ds^+\to \eta\pi^+$ decay 
 and the inclusive $\ds$ candidate is correctly reconstructed ($\ds\to \eta\pi$ decays); $\mu^+$ candidate originates from the $\ds\to\taunu\to \mu\nu_{\tau}\nu_{\mu}$ decay 
 and the inclusive $\ds$ candidate is correctly reconstructed ($\ds\to \taunu$); and all other candidates (combinatorial background). 
The $\mmiss^2(\dtagkx\gamma \mu)$ distributions is parameterized as:
\begin{eqnarray}
 {\cal F}(\mmiss^2) & = & N_{\rm sig}\cdot \sum_{i=1}^{3}{\cal G}(\mmiss^2,m^2_0,s_i\sigma^{MC}_i) + N_{\eta \pi}\sum_{i=1}^{2}{\cal G}(\mmiss^2,m^2_{\eta},\sigma_i^{MC})\nonumber\\
 & & + N_{K^0 K}{\cal BG}(\mmiss^2,m_0^{K^0K},\sigma_{BG}^{\rm left},\sigma_{BG}^{\rm right})\nonumber\\
 & & + N_{\tau}{ Exp}(t_{\tau}) + N_{\rm comb}{Exp}(t_{\rm comb}),
\end{eqnarray}
where the signal and the peaking background $\ds\to \eta \pi$ 
are parametrized using the sum of 3 and 2 Gaussian functions, respectively. All the parameters of the latter peaking background are fixed to values determined
on MC sample. In case of the signal all the shape parameters are fixed, except the mean ($m^2_0$). The common resolution scaling factor 
of the core and the second Gaussian $(s_1=s_2=s)$ is constrained (not fixed) to the value obtained on the $\ds\to K^0K$ and $\ds\to\eta\pi$ samples. 
The width of the third Gaussian (describing the outliers) is fixed. 
The $K^0K$ candidates are parametrized with a Bifurcated Gaussian function, where all parameters
are fixed to the values determined on MC sample. The $\ds\to\taunu$ and combinatorial backgrounds are parametrized with a exponential function, 
where the shape parameter of the combinatorial background is determined with the fit to the $\mmiss^2(\dtagkx\gamma \mu)$ distribution for candidates in the 
$\mmiss(\dtagkx\gamma)<1.95$ GeV sideband region and it is fixed to MC determined value for $\taunu$ decays. 
Yields of all event categories are free parameters of the fit, except for the
$\eta\pi$ and $K^0K$, which are constrained to expected values based on their known $\br$s and MC determined efficiencies. 
An important peaking background could arise from the $\ds^+\to \pi^+\pi^0$ 
decays, which would peak near the signal $\dsmunu$ decays (at $m^2_{\pi^0}=0.018$~GeV$^2$). 
These decays were not observed so far and based on the 
upper limit of $\br(\ds^+\to\pi^+\pi^0)<3.4\times 10^{-4}$ at 90\% C.L. set by CLEO-c \cite{Mendez:2009aa} 
their contribution is estimated to be less than 1 event. Their contribution is therefore negligible and is not taken into account.

\begin{figure}[t]
 \centering
 \includegraphics[width=1\textwidth]{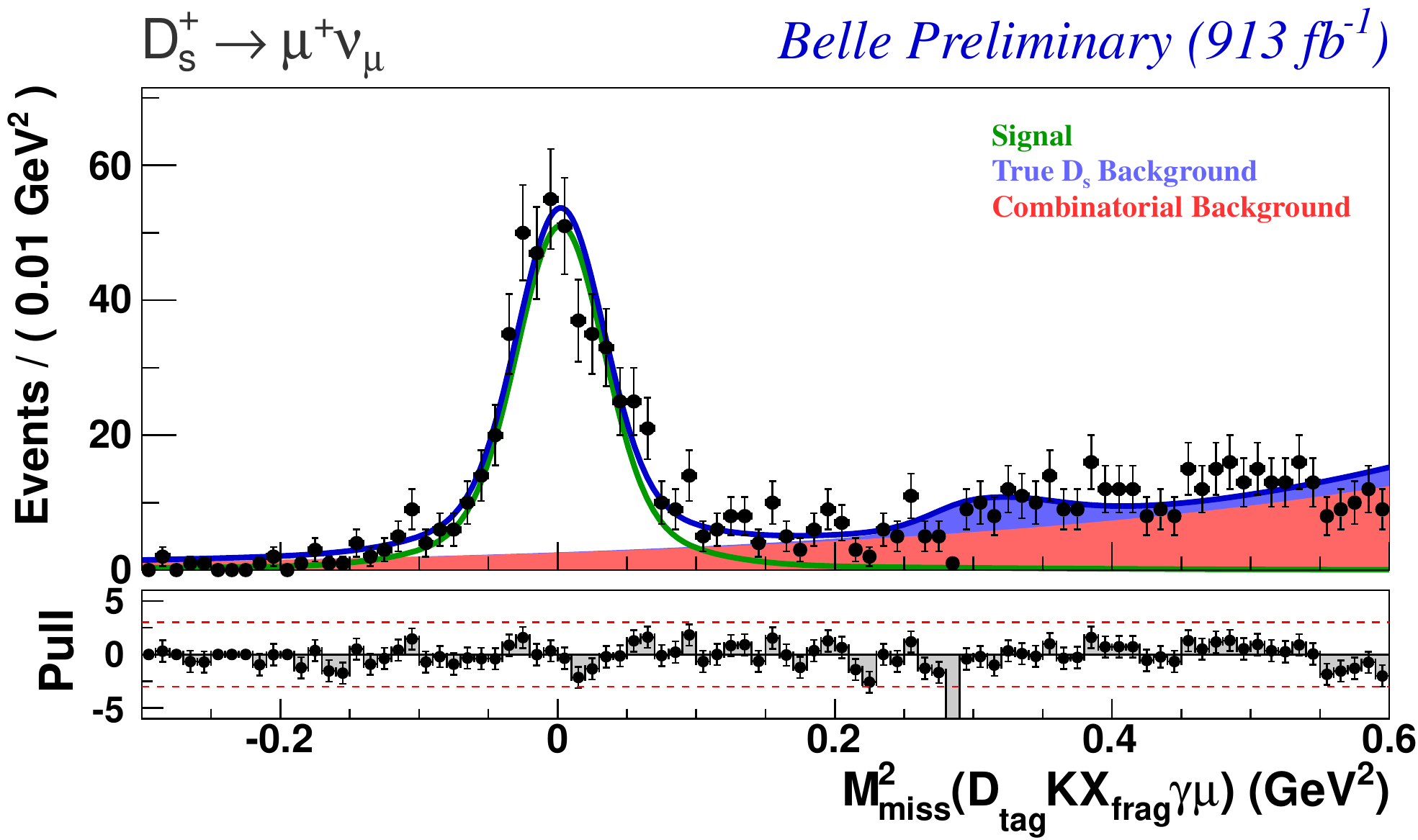} 
 \caption{The $\mmiss^2(\dtagkx\gamma \mu)$ distribution of exclusively reconstructed $\dsmunu$
 decays within the inclusive $\ds$ sample fit results superimposed (solid blue line). The solid green line shows the contribution
 of signal, while the contributions of combinatorial background and background other $\ds$ decays are indicated by the shaded red and blue
 areas, respectively.}
 \label{figs:munu:dataresults}
\end{figure}

The distribution of $\mmiss(\dtagkx\gamma\mu)$ with superimposed fit result is shown in Fig.\ref{figs:munu:dataresults}. 
The number of reconstructed $\ds\to\mu\nu$ decays is found to be
\begin{equation}
   N(\dsmunu)=489\pm26,
  \label{eq:dsmunu:yieldDA}
\end{equation}
where the error is statistical only. 

\subsection{\boldmath $\ds^+\to \tau^+\nu_{\tau}$}  
\label{sec:dstaunu}
The reconstruction of $\ds^+\to \tau^+\nu_{\tau}$ is performed by requiring one charged track to be present
in rest of the event and is identified as an electron, muon or a pion (denoted as $\ds\to\tau(X)\nu$ where $X=e,~\mu,~\pi$) indicating subsequent decay of $\tau$ lepton to $e\nu_e\nu_{\tau}$, 
$\mu\nu_{\mu}\nu_{\tau}$ or $\pi\nu_{\tau}$. Due to the multiple neutrinos in the final state these decays do not peak in the
missing mass squared distribution:
\[
 \mmiss^2(\dtagkx\gamma X) = p_{\rm miss}^2(\dtagkx\gamma X),
\]
where the missing 4-momentum is given by
\[
 p_{\rm miss}(\dtagkx\gamma X)= p_{e^+} + p_{e^-} - p_{\dtag} - p_{K} - p_{\xfrag} - p_{\gamma} - p_{X}.
\]

The background in $\ds\to\tau(\pi)\nu$ sample is much larger than in the leptonic modes which however can be significantly 
reduced by requiring the missing momentum of the event, $p_{\rm miss}(\dtagkx\gamma \pi)$, to be larger than 1.2 GeV. Background in 
this sample can be further reduced by requiring $0.0<\mmiss^2(\dtagkx\gamma \pi)<0.6$~GeV$^2$\footnote{Due to the lack of phase space and the 
fact that $\ds\to\taunu\to\pi\nu_{\tau}$ decays have only two neutrinos in the final state these decays populate relatively narrow region 
in $\mmiss^2(\dtagkx\gamma \pi)$.}. In case of $\ds\to\taunu\to\ell\nu_{\tau}\nu_{\tau}\nu_{\ell}$ we require $\mmiss^2(\dtagkx\gamma \ell)>0.3$ GeV$^2$ 
in order to veto contribution of $\ds^+\to\ell^+\nu_{\ell}$ decays. 

The signal yield of $\ds\to \tau(X)\nu_{\tau}$ decays is extracted from the fits to the $E_{ECL}$ distributions. The candidates are divided into
three categories: $X$ originates from the $\ds\to \tau(X)\nu_{\tau}$ decay and the inclusive $\ds$ candidate is correctly reconstructed (signal);
small contribution from other $\tau$ decays (cross-feed) is also considered as signal; $X$ originates from a $\ds$ decay such as semileptonic 
$h\ell\nu$ decays (when $X=\ell$) or hadronic $\ds$ decay modes (when $X=\pi$) and inclusive $\ds$ candidate is correctly reconstructed 
($\ds\to f$ background); and all other candidates (combinatorial background). The $E_{ECL}$ distribution is parametrized as:
\begin{eqnarray}
 {\cal F}(\eecl) & = & N_{\tau(X)}\cdot{\cal H}_{\rm \tau(X)}(\eecl)\nonumber\\
 & & + \sum_f N_{\ds\to f} \cdot{\cal H}_{\ds\to f}(\eecl) + N_{\rm comb}\cdot{\cal H}_{\rm comb}(\eecl),
\end{eqnarray}
where all components are described with non-parametric histogram PDF ($\cal H$) taken from simulated samples. 

\begin{figure}[t]
 \centering
 \includegraphics[width=0.49\textwidth]{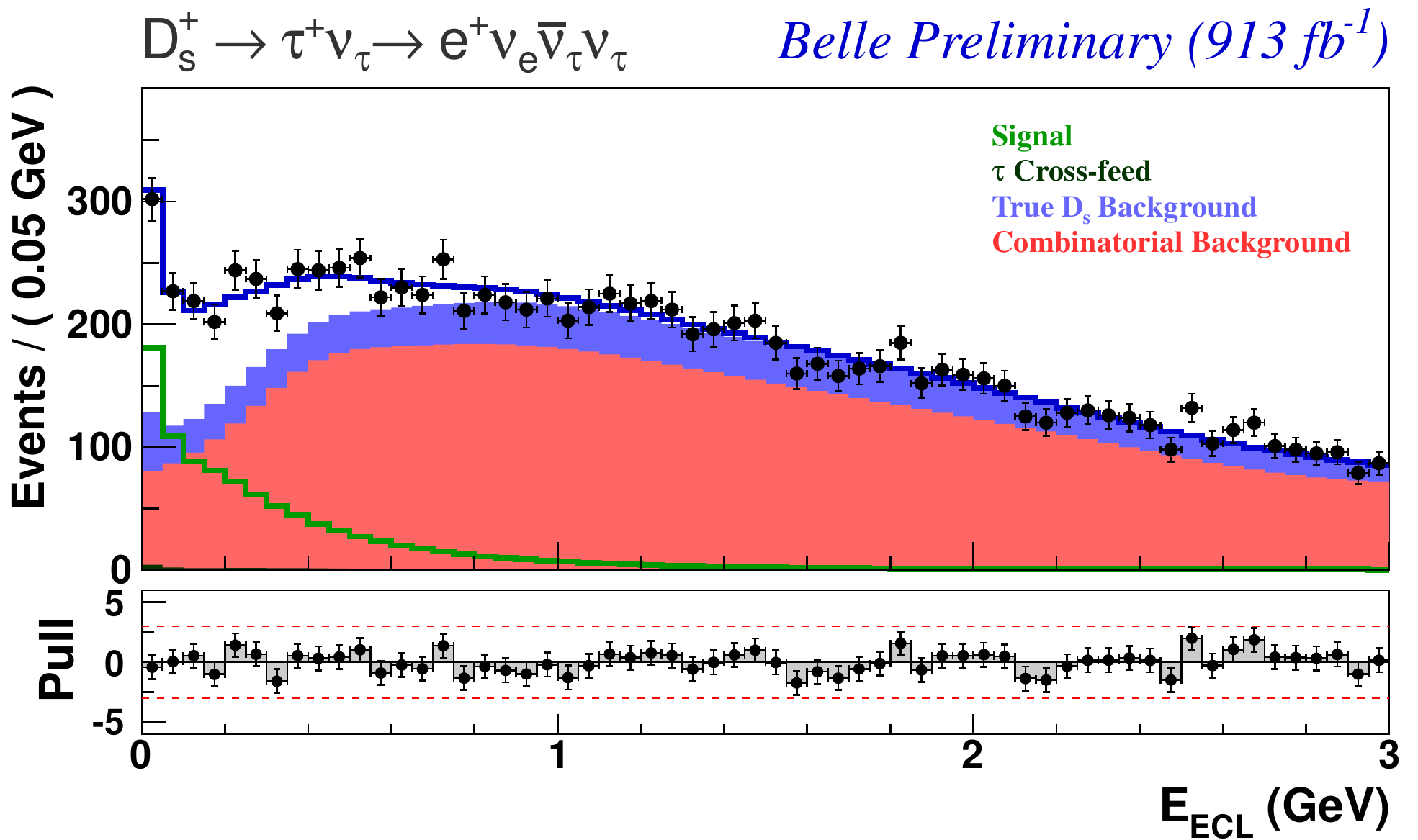}
 \includegraphics[width=0.49\textwidth]{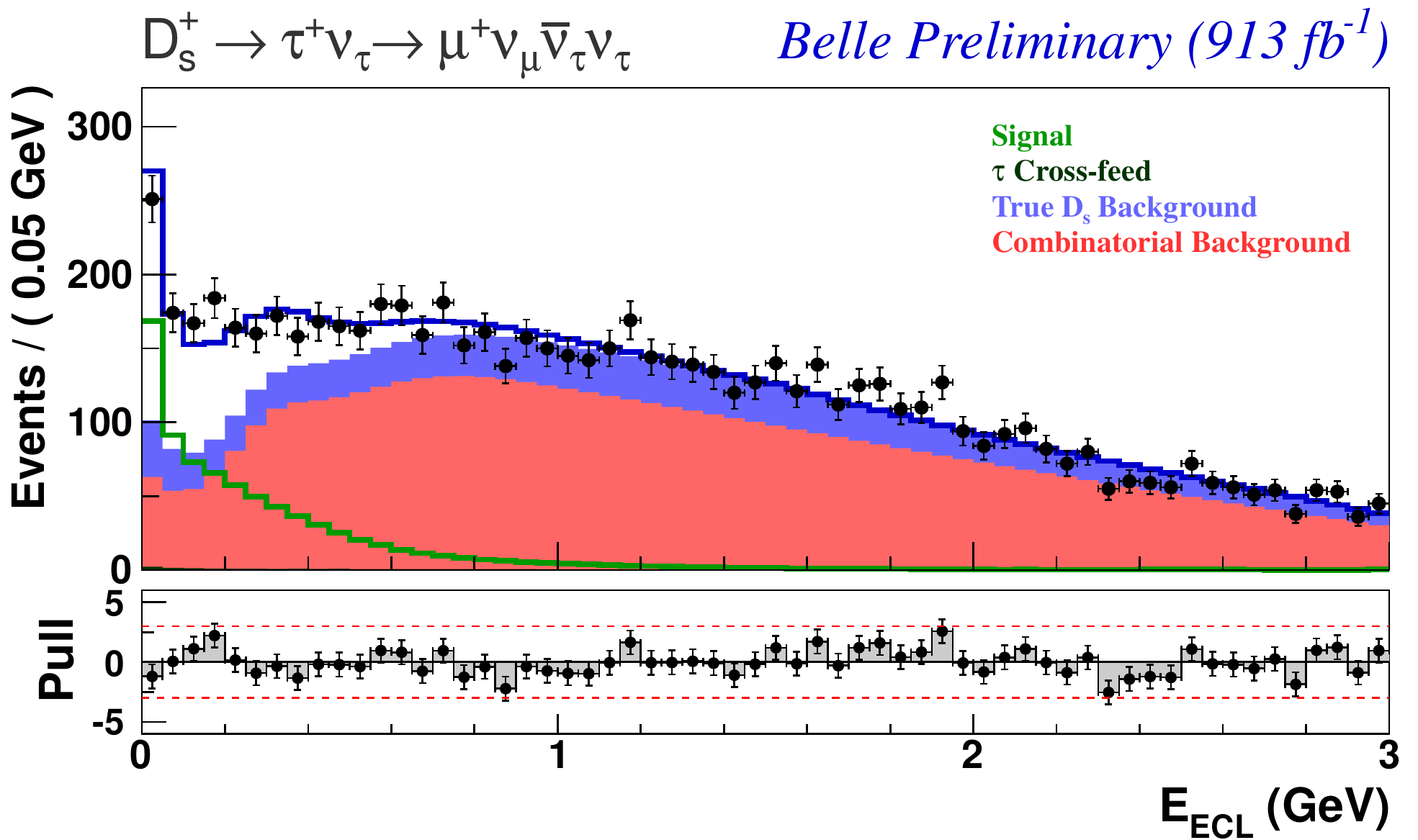}\\  
 \includegraphics[width=0.49\textwidth]{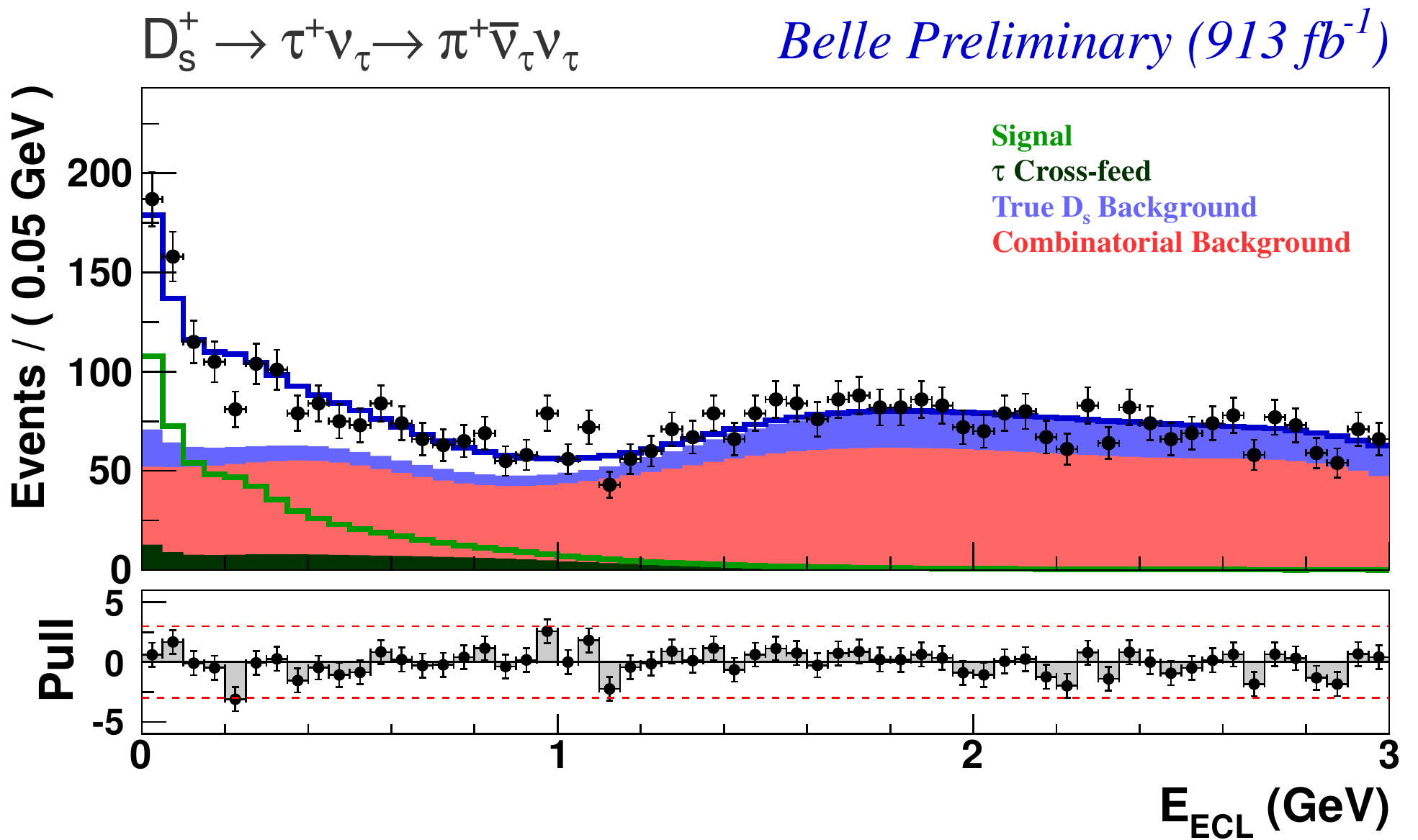}\\ 
 \caption{The $\eecl$ distribution of exclusively reconstructed $\ds\to\tauenu$ (top left), $\ds\to\taumunu$ (top right), and $\ds\to\taupinu$ (bottom) 
 decays within the inclusive $\ds$ sample with fit results superimposed. The solid green lines show the contribution of signal and the
  contribution of $\tau$ cross-feeds are indicated by the full dark green histograms. The contributions of combinatorial background and 
   background from other $\ds\to f$ decays is indicated by the shaded red and blue areas, respectively.}
 \label{figs:taunu:dataresults}
\end{figure}

The distributions of $E_{\rm ECL}$ for $\ds\to\tau(e)\nu$, $\ds\to\tau(\mu)\nu$, and $\ds\to\tau(\pi)\nu$ decays are shown in 
Fig.~\ref{figs:taunu:dataresults} with fit results superimposed. The number of reconstructed signal decays are found to be:
\begin{eqnarray}
 N(\ds\to\tauenu)&=&952\pm59,\\
 N(\ds\to\taumunu)&=&758\pm48\\
 N(\ds\to\taupinu)&=&496\pm35,
\end{eqnarray}
where the errors are statistical only. 

\section{Results and Systematics}
\begin{table}[t]
 \begin{tabular}{lccc}\hline\hline
  $\ds^+$ Decay Mode	& Signal Yield				& $f_{\rm bias}\cdot\varepsilon(\ds\to f|{\rm incl.}~\ds)$	& $\br$ [\%] \\
  \hline\hline
  $K^-K^+\pi^+$		& $4094\pm123$				& $0.8578\pm0.0056$			& $5.06 \pm 0.15 \pm 0.19$\\
  $\overline{K}{}^0K^+$	& $1943\pm82\phantom{0}$		& $0.7254\pm0.0054$			& $2.84 \pm 0.12 \pm 0.08$\\
  $\eta\pi^+$		& $\phantom{0}773\pm58\phantom{0}$	& $0.4577\pm0.0064$			& $1.79 \pm 0.14 \pm 0.05$\\\hline
  $\munu$		& $\phantom{0}489\pm26\phantom{0}$	& $0.9813\pm0.0175$			& $0.528\pm0.028\pm0.019$\\\hline
  $\taunu$ ($e$ mode)	& $\phantom{0}952\pm59\phantom{0}$	& $1.0531\pm0.0135$			& $5.37\pm0.33{}^{+0.35}_{-0.30}$\\
  $\taunu$ ($\mu$ mode)	& $\phantom{0}758\pm48\phantom{0}$	& $0.7849\pm0.0118$			& $5.88\pm0.37{}^{+0.34}_{-0.58}$\\
  $\taunu$ ($\pi$ mode)	& $\phantom{0}496\pm35\phantom{0}$	& $0.8072\pm0.0152$			& $5.96\pm0.42{}^{+0.45}_{-0.39}$\\\hline
  $\taunu$ (Combined)& \multicolumn{2}{c}{}								& $5.70\pm0.21{}^{+0.31}_{-0.30}$\\\hline\hline
 \end{tabular}
 \caption{Signal yields, tag bias corrected efficiencies and measured branching fractions for all five studied $\ds^+$ decay modes. 
 The first uncertainty is statistical and the second is systematic.}
 \label{tab:results}
\end{table}

From the extracted signal yields of studied $\ds$ decay modes we determine their absolute branching fractions using Eq.~\ref{eq:abs_br_ideal}.
They are summarized in Table~\ref{tab:results} and are found to be consistent with previous measurements performed by CLEO and Babar
collaborations. 

Systematic errors for the measured branching fractions
are associated with the uncertainties in the signal yields,
the efficiencies, and the number of inclusively reconstructed $\ds$. 

The systematic error\footnote{All systematic errors are given as relative uncertainties.} related to the normalization is assigned to 
be 1.95\% (see Eq.~\ref{eq:inclds:yield}) and is common for all studied $\ds\to f$ decays. 
As described in Sec. \ref{sec:brandyields} the $\fbias$ factor is corrected by the 
${\overline{\varepsilon}{}_{\ds}^{\rm inc.}|_{\rm DATA}}/{\overline{\varepsilon}{}_{\ds}^{\rm inc.}|_{\rm MC}}$ ratio and the uncertainty
on the correction (1.37\%) is assigned as systematic error referred to as tag bias and is common for all $\ds\to f$ decays. 
The systematic errors in the $\ds\to f$ reconstruction efficiencies arise from the uncertainty in tracking efficiency (0.35\% per 
reconstructed charged track in the final state of  $\ds\to f$ decay), particle identification efficiency, branching fractions of 
$\tau$ decays, and MC statistics. All of these depend on the studied $\ds\to f$ decay.
The systematic error from MC statistics of the histogram PDFs is evaluated by varying the content of each bin by its statistical uncertainty
in case of $\ds\to\taunu$ signal yields. 
The shape parameters of combinatorial background are varied within their uncertainties and the differences with respect to the nominal fits are
taken as the systematic uncertainty in case of $\dsmunu$, $\ds\to K^0K^+$, and $\ds\to\eta\pi$ decays. To estimate the systematic error
due to the possible signal $\eecl$ shape difference between MC and data, the ratio of data to MC for the
background subtracted $\eecl$ histograms of the $\ds\to KK\pi$ and $\ds\to K^0K^+$ samples is fitted with a fourth-order 
polynomial and the signal $\eecl$ PDF is modified within the fitted errors. Similarly we estimate the systematic error
due to the possible combinatorial $\eecl$ shape between MC and data by fitting the ratio of $\eecl$ histograms of $\ds\to\taunu$ candidates
populating the $\mmiss(\dtagkx\gamma)$ sideband region (defined as $\mmiss(\dtagkx\gamma)<1.94$~GeV or $\mmiss(\dtagkx\gamma)>2.00$~GeV) with
a linear function and modifying the combinatorial $\eecl$ PDF within the fit errors. 
The uncertainties for the branching fractions of $\ds$ decays the true $\ds$ background categories that are fixed in the fits are 
estimated by changing the branching fractions by their experimental errors \cite{PDG}. In case of $\ds\to\tauenu$ and
 $\ds\to\taumunu$ decays the largest contribution to the systematic uncertainty originates from $\ds\to K^0\ell\nu$ decays and in case of
 $\ds\to\taupinu$ from $\ds\to\eta\pi$ and $\rho K$ decays. The $\tau$ cross-feed is fixed relative to the signal contribution in the nominal fit to 
the $\eecl$ distributions of $\ds\to\tau\nu$ samples. The ratios
 are varied within their uncertainties and the fits are repeated and the differences 
from the nominal fits are taken as the systematic uncertainties.

The total systematic error is calculated by summing the above uncertainties in quadrature. The estimated systematic errors are summarized
in Tables \ref{tab:syst:hadr} and \ref{tab:sytematics:ellnu} for the hadronic and leptonic $\ds$ decay modes, respectively.
\begin{table}[t]
 \centering
 \begin{tabular}{lccc}\hline\hline
  Source        		& $K^-K^+\pi^+$ [\%]	& $\overline{K}{}^0K^+$ [\%]	& $\eta\pi^+$ [\%]
  \\\hline\hline
  Statistical   		& 3.00			& 4.22				& 7.50\\ 
  \hline
  Normalization 		& 1.95			& 1.95				& 1.95\\
  Tag bias      		& 1.37 			& 1.37				& 1.37\\
  Tracking      		& 1.05 			& 0.35				& 0.35\\
  Efficiency    		& 0.65 			& 0.74				& 1.40\\
  PID           		& 2.57 			& 0.82				& 1.08\\
  Signal PDF    		& 0.82 			& --				& --\\
  True $\ds$ background		& --			& 0.56				& 0.65\\
  \hline\hline
  Total syst.   		& 3.81 			& 2.71 				& 3.06\\ 
  \hline
  Stat. + Syst. 		& 4.85 			& 5.02 				& 8.10\\ 
  \hline\hline
 \end{tabular}\\
\caption{Summary of systematic uncertainties for the branching fraction measurement of hadronic $\ds$ decays.}
\label{tab:syst:hadr}
\end{table}

\begin{table}[t!]
 \centering
 \renewcommand{\arraystretch}{1.2}
 \begin{tabular}{lccccc} \hline\hline
  Source        		& $\mu\nu$ [\%]		& $\tau(e)\nu$ [\%]	& $\tau(\mu)\nu$ [\%]	& $\tau(\pi)\nu$ [\%]	& $\tau\nu$ [\%]  \\\hline\hline
  Statistical			& $\pm 5.32$		& $\pm 6.18$		& $\pm 6.33$		& $\pm 7.04$		& $\pm 3.75$\\\hline
  Normalization 		& $\pm 1.95$		& $\pm 1.95$		& $\pm 1.95$		& $\pm 1.95$		& $\pm 1.95$\\
  Tag bias 			& $\pm 1.37$		& $\pm 1.37$		& $\pm 1.37$		& $\pm 1.37$		& $\pm 1.37$\\
  Efficiency			& $\pm 1.78$		& $\pm 1.28$		& $\pm 1.51$		& $\pm 1.88$		& $\pm 0.84$\\
  Tracking 			& $\pm 0.35$		& $\pm 0.35$		& $\pm 0.35$		& $\pm 0.35$		& $\pm 0.35$\\
  PID 				& $\pm 1.96$		& $\pm 2.03$ 		& $\pm 1.93$		& $\pm 0.88$		& $\pm 1.70$\\
  Signal PDF			& --			& $+3.46$		& $+1.96$		& $+3.43$		& $+2.95$\\
  Comb. bkg. PDF		& $\pm 0.02$		& $+0.11$		& $-8.31$		& $+ 0.92$		& $-2.54$\\
  PDF stat.			& --			& $\pm 2.16$		& $\pm 2.19$		& $\pm 3.05$		& $\pm 1.44$\\
  True $D_s$ background		& $\pm 0.82$		& $\pm 3.88$		& $\pm 3.56$		& $\pm 3.15$		& $\pm 2.84$\\
  $\tau$ cross-feed		& --			& $\pm 0.36$		& $\pm 0.24$		& $\pm 3.71$		& $\pm 0.94$\\
  ${\cal B}(\tau\to X)$		& --			& $\pm 0.22$		& $\pm 0.23$		& $\pm 0.64$		& $\pm 0.19$\\\hline\hline
  Total syst. 			& $\pm 3.67$		& $^{+6.59}_{-5.61}$	& $^{+5.76}_{-9.92}$ 	& $^{+7.49}_{-6.60}$	& $^{+5.40}_{-5.19}$\\ \hline
  Stat. + Syst. 		& $\pm 6.46$		& $^{+9.03}_{-8.35}$	& $^{+8.56}_{-11.8}$ 	& $^{+10.3}_{-9.65}$	& $^{+6.57}_{-6.40}$\\ \hline\hline
\end{tabular} 
\caption{Summary of systematic uncertainties for the branching fraction measurement of leptonic $\ds$ decays.}
\label{tab:sytematics:ellnu}
\end{table}

\section{\boldmath Extraction of $\fds$ and Conclusions}
The value of $\fds$ is determined from measured branching fractions of leptonic $\ds$ decays. Inverting Eq. \ref{eq:brleptonic_sm} yields
\[
\fds=\frac{1}{G_Fm_{\ell}\left( 1-\frac{m_{\ell}^2}{M_{\ds}^2}\right)\left|V_{cs}\right|} \sqrt{\frac{8\pi\br(\ds\to\ell\nu_{\ell})}{M_{\ds}\tau_{\ds}}}.
\]
The external inputs necessary in the extraction of $\fds$ from the measured $\br$s are given in Table \ref{tab:fds:externalinput}. 
The $|V_{cs}|$ is obtained from the very well measured $|V_{ud}|=0.97425(22)$ and $|V_{cb}|=0.04$ from an average of 
exclusive and inclusive semileptonic B decay results
as discussed in Ref. \cite{vcb} by using the following relation:
\[
 |V_{cs}|=|V_{ud}|-\frac{1}{2}|V_{cb}|^2.
\]
The external inputs are all very precisely measured and do not introduce additional uncertainties except the $\ds$ lifetime, $\tau_{\ds}$, which
introduces an 0.70\% relative uncertainty on $\fds$.
\begin{table}[t]
 \centering
 \begin{tabular}{lr}\hline\hline
  Quantity & Value \\\hline\hline
  $M_{\ds}$  & 1.96847(33) GeV \\
  $m_{\tau}$ & 1.77682(16) GeV \\
  $m_{\mu}$  & 0.105658367(9) GeV \\
  $\tau_{\ds}$ & 0.500(7) ps\\
  $G_F$ & $1.16637(1)\times 10^{-5}$ GeV$^{-2}$\\
  $|V_{cs}|$ & 0.97345(22)\\\hline\hline
 \end{tabular}
 \caption{Numerical values of external parameters used in extraction of $\fds$.}
 \label{tab:fds:externalinput}
\end{table}
Table \ref{tab:fds:belle} summarizes the obtained values of $\fds$ using the $\ds^+\to\mu^+\nu_{\mu}$ and $\ds^+\to\tau^+\nu_{\tau}$ decays. An error-weighted 
average of $\fds$ is found to be
\[
 \fds = (255.0\pm4.2(\rm stat.)\pm4.7(\rm syst.)\pm1.8(\tau_{D_s}))~\mbox{MeV},
\]
where the correlation of the systematic uncertainties between the $\mu\nu$ and $\tau\nu$ have been taken into account. 
This is the most precise measurement of $\fds$ up to date. 
\begin{table}[t!]
 \centering
 \begin{tabular}{lc}
  $D_s\to \ell\nu$  & $f_{D_s}$ [MeV] \\ \hline\hline
  $\mu\nu$          & $249.0\pm6.6({\rm stat.})\pm4.6({\rm syst.})\pm1.7(\tau_{D_s})$ \\
  $\tau\nu$         & $261.9\pm4.9({\rm stat.})\pm7.0({\rm syst.})\pm1.8(\tau_{D_s})$ \\\hline
  Combination       &$255.0\pm4.2({\rm stat.})\pm4.7({\rm syst.})\pm1.8(\tau_{D_s})$ \\\hline\hline
 \end{tabular}
 \caption{Measured values of $\fds$ in $\mu\nu$ and $\tau\nu$ decay modes and the combination of the two.}
 \label{tab:fds:belle}
\end{table}

\Acknowledgements
We thank the KEKB group for the excellent operation of the
accelerator; the KEK cryogenics group for the efficient
operation of the solenoid; and the KEK computer group,
the National Institute of Informatics, and the 
PNNL/EMSL computing group for valuable computing
and SINET4 network support.  We acknowledge support from
the Ministry of Education, Culture, Sports, Science, and
Technology (MEXT) of Japan, the Japan Society for the 
Promotion of Science (JSPS), and the Tau-Lepton Physics 
Research Center of Nagoya University; 
the Australian Research Council and the Australian 
Department of Industry, Innovation, Science and Research;
the National Natural Science Foundation of China under
contract No.~10575109, 10775142, 10875115 and 10825524; 
the Ministry of Education, Youth and Sports of the Czech 
Republic under contract No.~LA10033 and MSM0021620859;
the Department of Science and Technology of India; 
the Istituto Nazionale di Fisica Nucleare of Italy; 
he BK21 and WCU program of the Ministry Education Science and
Technology, National Research Foundation of Korea Grant No.\ 
2010-0021174, 2011-0029457, 2012-0008143, 2012R1A1A2008330,
BRL program under NRF Grant No. KRF-2011-0020333,
and GSDC of the Korea Institute of Science and Technology Information;
the Polish Ministry of Science and Higher Education and 
the National Science Center;
the Ministry of Education and Science of the Russian
Federation and the Russian Federal Agency for Atomic Energy;
the Slovenian Research Agency;
the Basque Foundation for Science (IKERBASQUE) and the UPV/EHU under 
program UFI 11/55;
the Swiss National Science Foundation; the National Science Council
and the Ministry of Education of Taiwan; and the U.S.\
Department of Energy and the National Science Foundation.
This work is supported by a Grant-in-Aid from MEXT for 
Science Research in a Priority Area (``New Development of 
Flavor Physics''), and from JSPS for Creative Scientific 
Research (``Evolution of Tau-lepton Physics'').

\end{document}

%% file: econfmacros.tex
\def\beq{\begin{equation}}
\def\eeq#1{\label{#1}\end{equation}}
\def\eeqn{\end{equation}}

\def\beqa{\begin{eqnarray}}
\def\eeqa#1{\label{#1}\end{eqnarray}}
\def\eeqan{\end{eqnarray}}

\let\bar=\overbar

\def\etal{{\it et al.}}

\def\Dslash{\not{\hbox{\kern-4pt $D$}}}
\def\dslash{\not{\hbox{\kern-2pt $\del$}}}

\def\msb{{\bar{\ssstyle M \kern -1pt S}}}